\documentclass[journal]{IEEEtai}

\ifCLASSINFOpdf
   \usepackage[pdftex]{graphicx}
   \graphicspath{{../pdf/}{../jpeg/}}
   \DeclareGraphicsExtensions{.pdf,.jpeg,.png}
\else
   \usepackage[dvips]{graphicx}
   \graphicspath{{../eps/}}
   \DeclareGraphicsExtensions{.eps}
\fi

\usepackage{color,array}
\usepackage{amssymb,amsmath,bm}
\usepackage{textcomp}
\usepackage{multirow}
\usepackage[export]{adjustbox}
\usepackage{hyperref}
\usepackage{subfig}
\usepackage{url}
\usepackage{harmony}
\usepackage[boxed]{algorithm2e}

\DeclareMathOperator{\E}{\mathbb{E}}
\hypersetup{colorlinks=false, urlcolor=blue}

\begin{document}

\title{Automatic Neural Lyrics and Melody Composition}

\author{Gurunath Reddy M, Yi Yu, Florian Harsco\"et, Simon Canales, Suhua Tang
\thanks{Gurunath Reddy M is with Indian Institute of Technology, Kharagpur, India (email: mgurunathreddy@sit.iitkgp.ernet.in)}
\thanks{Yi Yu is with National Institute of Informatics, Japan, (email: yiyu@nii.ac.jp)}
\thanks{Florian Harsco\"et is with ISIMA, France, (email: florian.harscoet@etu.uca.fr)}
\thanks{Suhua Tang is with The University of Electro-Communications, Japan, (email: shtang@uec.ac.jp)}
\thanks{Simon Canales is with EPFL, Switzerland, (email: simon.canales@epfl.ch)}
\thanks{This paragraph will include the Associate Editor who handled your paper.}}


\maketitle

\begin{abstract}
In this paper, we propose a technique to address the most challenging aspect of algorithmic songwriting process, which enables the human community to discover original lyrics, and melodies suitable for the generated lyrics. The proposed songwriting system, Automatic Neural Lyrics and Melody Composition (AutoNLMC) is an attempt to make the whole process of songwriting automatic using artificial neural networks. Our lyric to vector (lyric2vec) model trained on a large set of lyric-melody pairs dataset parsed at syllable, word and sentence levels are large scale embedding models enable us to train data driven model such as recurrent neural networks for popular English songs. AutoNLMC is a encoder-decoder sequential recurrent neural network model consisting of a lyric generator, a lyric encoder and melody decoder trained end-to-end. AutoNLMC is designed to generate both lyrics and corresponding melody automatically for an amateur or a person without music knowledge. It can also take lyrics from professional lyric writer to generate matching melodies. The qualitative and quantitative evaluation measures revealed that the proposed method is indeed capable of generating original lyrics and corresponding melody for composing new songs.
\end{abstract}

\begin{IEEEImpStatement}
Recently, we can see a surge in growing research interest in automatically generating music after the successful application of deep neural networks for sequential modeling. Most of the works focus on either generating melodies or lyrics. However, to the best of our knowledge, there are no works which focus on automating the song creation process which has potential application in cinema and gaming industries, and educational pedagogy. The proposed AutoNLMC can be used by both amateur and professionals to create songs with both lyrics and melodies automatically. Further, generated lyric melody pairs can be used to synthesize songs using synthesizers which can be readily used in gaming and cinema applications. Also, using AutoNLMC one can discover new melodies for already existing lyrics without melodies which enables a faster song creation process.     
\end{IEEEImpStatement}


\section{Introduction}

\IEEEPARstart{M}{usic} composition is a human creative process that requires a wide range of strong musical knowledge and expertise to create soothing music which continues to remain in our heart forever. Given the vast majority of music lovers and the limited availability of professional music composers, there is a strong need for machines to assist human creativity. Recent advancement in the software based music creation technology helped the professional and amateur music creators to produce music with great joy and ease of production in masses to be consumed by the music consumers with personal computers and hand-held devices. 
Though there exists a plenty of machine assistance to create high quality music with relative ease of production, the process of songwriting that is automatically generating lyrics, composing melody corresponding to the generated lyrics and synthesizing singing voice corresponding to the generated melody and lyrics remained as mutually exclusive tasks. Till date, the construction of novel/original songs is limited to the individuals who possess the following skills: the ability to create lyrics, compose melody and combine lyrics and melody to create a rational, relevant and soothing final complete songs~\cite{ackerman2017algorithmic}. 

\begin{figure*}[htbp]
    \centering
    \includegraphics[width=\textwidth]{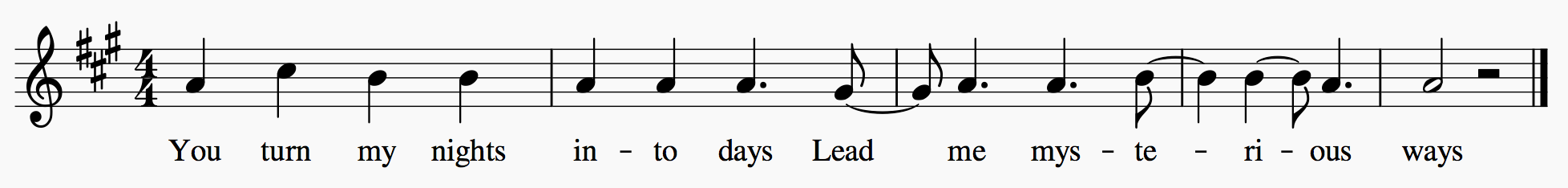}
    \caption{Example music score}
    \label{fig:score}
\end{figure*}
         
In literature, we can find considerable amount of research work published on automatic music generation (without conditioning on lyrics). Early machine assisted music generation is mostly based on music theory and expert domain knowledge to create novel works. With the advent of data driven approaches and exploded public music collections in the internet, data driven methods such as Hidden Markov models~\cite{pachet2011markov}, graphic models~\cite{pachet2017sampling} and deep learning models~\cite{waite2016generating, chu2016song, mogren2016c, dong2018musegan, yu2019deep, yu2019conditional} showed a potential for music creation. Though there exists substantial amount of research on unconditional music generation, there exists considerably less amount of work done so far on generating melody from lyrics given in the form of text, which we call conditional melody/song generation from lyrics. The primary reasons for substantially less research on conditional melody generation can be attributed to i) the non-availability of the direct source for lyrics-melody pair dataset to train the data driven models, ii) a lyrics composition can have multiple melodic representations, which makes it hard to learn the correlation between the lyrics and melodies, and iii) it is hard to evaluate the generated melodies by objective measures.

This paper focuses on the most challenging aspect of algorithmic songwriting process which enables the human community to discover original lyrics, and  melodies suitable for the generated lyrics. To the best our knowledge, the proposed AutoNLMC is the first attempt to make the whole process of songwriting automatic using artificial neural networks. We also present the lyrics to vector model which is trained on a large dataset of popular English songs to obtain the dense representation of lyrics at syllables, words and sentence levels. The proposed AutoNLMC is an attention based encoder-decoder sequential recurrent neural network model consists of a lyric generator, lyric encoder and melody decoders trained end-to-end. We train several encoder-decoder models on various dense representations of the lyric tokens to learn the correlation between lyrics and corresponding melodies. Further, we prove the importance of dense representation of lyrics by various qualitative and quantitative measures. AutoNLMC is designed in such a way that it can generate both lyrics and corresponding melodies automatically for an amateur or a person without music knowledge by accepting a small piece of initial seed lyrics as input. It can also take lyrics from professional lyrics writer to generate the matching meaningful melodies.

\section{Related Work}
\label{sec:rel_work}

Here, we briefly present closely related work on lyrics-conditional music generation frameworks developed by researchers in  past years. Orpheus~\cite{fukayama2010automatic} is a dynamic programming based melody composition algorithm for Japanese lyrics. Orpheus is designed as an optimal melody search problem for the given lyrics under the prosodic constraints of the Japanese lyrics. The authors design two individual models such as rhythm and probabilistic pitch inference model to generate melodies from the given lyrics. A Finish song generating system called Sucus-Apparatusf is developed by Toivanen et al.~\cite{toivanen2013automatical}. Sucus-Apparatusf is designed to randomly choose rhythm from the rhythm patterns actually found in the Finish art songs. Further, a second order Markov model is designed to generate the chord progression for the given lyrics. The pitches are generated from the joint probabilistic distribution of previously generated notes and the chords. In~\cite{monteith2012automatic}, the authors propose a system for automatically generating melodic accompaniments from a given lyrical text. The system is designed to generate the pitches modeled as n-gram models from the melodies of songs with similar style. Further, the rhythm for the melodic accompaniment is derived from the cadence information present in the text. The proposed method generated hundreds of melodies by giving random options driven by set of rules, followed by selecting among the generated options with an objective measure that incorporates expert music knowledge. Both melody and rhythm are generated by the same process for a given lyrics. Rhythm suggestion from lyrics by using set of rules is studied by Nichols~\cite{nichols2009lyric} while Oliveira~\cite{oliveira2015tra} proposed the inverse process of lyrics generation from the rhythm of melody. 
ALYSIA~\cite{ackerman2017algorithmic} is the first fully data driven model based on random forests to generate melody from the lyrics. ALYSIA is trained on a large set of features manually extracted from Music-XML files. ALYSIA is designed to suggest multiple melodies as output for a given lyrical piece thus giving user the ability to choose more pleasing melody for the given lyrics. ALYSIA consists of two independent melody prediction models to predict duration of the note and scale of the note independently. An encoder-decoder based RNN sequential model for lyrics-conditional melody generation for Chinese pop songs is presented in~\cite{bao2018neural}. The sequential model called Songwriter consists of two encoders and one hierarchical decoder. The encoders are designed to encode the lyric syllables and context melody of the prior generated melody. The hierarchical decoder is designed to decode the note attributes such as pitch, duration and syllable-note alignment labels since most of the syllables in Chinese songs had more than one note.     


\section{Music Background}
\label{sec:music_background}

Melody can be defined as a sequence of meaningful musical notes and musical rests. Let $N$ be the number of notes of a given melody. Each note of the melody has two attributes: 
\begin{enumerate}
    \item its pitch -- called note or tone in common musical language -- which is represented by a letter followed by some number (e.g. C4) (and eventually a $\flat$ (flat) or $\sharp$ (sharp) symbol (e.g. F$\sharp$5));
    \item its duration, which depends on the note type.
\end{enumerate}

The common unit used for the duration is the {beat}. In most modern Western music, four beats form a {measure}. A note which lasts for one beat (or a quarter measure) is called a quarter-note. Accordingly other notes can be created: sixteenth, eighth, etc. Then, between successive notes there is either a rest or no rest. Again, the rest duration unit is the beat.

Therefore, a melody of length $N$ can be defined as $Z=\{\mathbf{z}_1,\mathbf{z}_2,\dots,\mathbf{z}_N\}$, where each $\mathbf{z}_i$'s are \{note pitch, note duration, rest duration\} triplets (which is the representation used for this work). The $i$-th rest duration value is the duration of the rest before the $i$-th note. A null rest duration means no rest before the note.

We can define lyrics as sequence of natural language tokens which can eventually make up song consisting of choruses and verses. The lyrics token representation adopted in our framework is a syllable. A syllable is a part of word or a single-syllable word. Each syllable is associated with melody attributes: pitch, duration and rest to from the lyric-melody pairs.

We can represent lyrics-melody pair graphically via music scores. Each syllable is associated with the corresponding note (see Figure \ref{fig:score}). The shape of the notes gives their duration, and their vertical position their pitch. The duration of a rest is given by its shape. Most of the time, melodies have perfect {scale consistency}, meaning that the pitches of the notes composing the melody all belong to the same {scale}. A scale is a subset of pitches which have properties such that they sound good when consecutively played.

\section{Probabilistic Modeling of AutoNLMC}
\label{sec:prob_def}

The proposed AutoNLMC is designed to first generate lyrics given a piece of seed lyrics in the form of text and then composes the melody for the generated lyrics sequentially one sentence at a time. 
We can model the AutoNLMC as a probabilistic model with two conditionally dependent components: a lyrics generator and a melody composer sequential model. Lyrics generator is modeled as a conditional distribution to predict the next lyrical token given the previous tokens of the lyric sequence modeled as $p(s_t | s_1, s_2, ..., s_{t-1})$. Here, the tokens are the sequence of syllables of the lyrics given by $S = \{s_1, s_2, . . ., s_T\}$. We can learn a probability distribution such as 

\begin{equation}
\label{eq:lyrics_generator}
p(S) = \prod_{t=1}^{T} p(s_t | s_1, s_2, ..., s_{t-1})
\end{equation}       

\noindent the learned probability distribution is sampled one token at every time to generate the full sequence lyrics.

We can model the Lyrics generator using recurrent neural network (RNN) to implicitly learn the conditional probability distribution of the lyrics generator. RNN can be trained to learn the probability distribution of a sequence to predict the next token of the sequence at time $t$. RNN is essentially an artificial neural network  consisting of a hidden state $h$, an output $y$, operates on a sequence of tokens to update the hidden states at each time step $t$ given by

\begin{equation}
\label{eq:hidden_state}
h_t = g(h_{t-1}, s_t)
\end{equation}          
       
\noindent where $g$ is a non-linear function mapping from input and previous state to the present state. The non-linear function $g$ can be a simple sigmoid function~\cite{cho2014learning} or it can be a more complex gated unit such as long short-term memory (LSTM)~\cite{hochreiter1997long}. Melody composer is modeled as a conditional distribution model over melody sequence on a lyrics sequence given by $p(m_1, m_2, .., m_{T} | s_1, s_2, ..., s_T)$ where $M = \{m_1, m_2, .., m_{T}\}$ is a melody sequence. The melody composer is modeled as a sequential encoder-decoder model~\cite{sutskever2014sequence}. The encoder is an RNN network which takes the lyrics tokens generated by lyrics generator model Eq.~\ref{eq:lyrics_generator} at each time step sequentially to produce the encoder hidden states at time $t$ defined in Eq.~\ref{eq:dec_hidden_state}. The encoder also generates a variable context vector $C_i$ which encodes the parts of the lyrics for which most attention to be paid during melody decoding. The decoder is another RNN trained to predict the melody token at time $t$ given the hidden state of the decoder RNN. The hidden state of the decoder is computed recursively by             

\begin{equation}
\label{eq:dec_hidden_state}
h_t = g(h_{t-1}, m_{t-1}, C_i)
\end{equation}          

\noindent The next token distribution of the decoder can be similarly defined as 

\begin{equation}
p(m_t | m_1, m_2, ...., m_{t-1}, C_i) = f(h_t, m_{t-1}, C_i)
\end{equation} 

\noindent The melody composer can be jointly trained to maximize the likelihood

\begin{equation}
\operatorname*{max}_\theta \frac{1}{N} \sum_{i=1}^{N} \log p_\theta(m_i | s_i)
\end{equation}

\noindent where $\theta$ is the model parameters and the pair $m_i, s_i$ is the input and output to the melody composer.

\begin{figure}[htbp]
        \centering
		\resizebox{8cm}{8cm}{\includegraphics{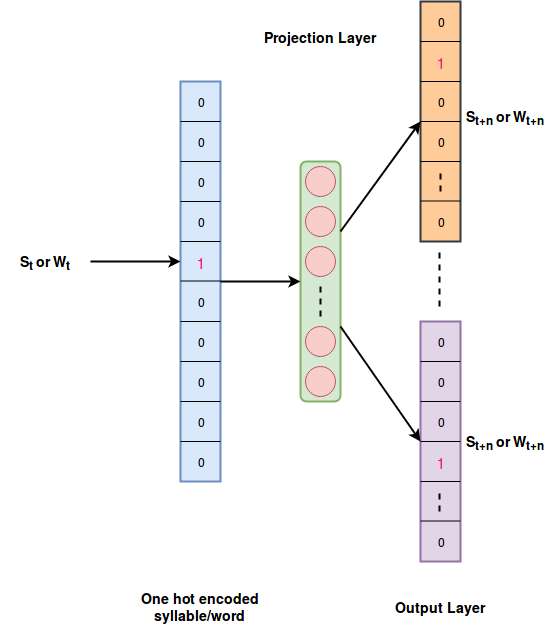}}  
        \caption{Skip-gram model to extract lyrics embedding.}
        \label{fig:skip_gram}
 \end{figure} 

\begin{figure*}[htbp]
        \centering
		\resizebox{14cm}{8cm}{\includegraphics{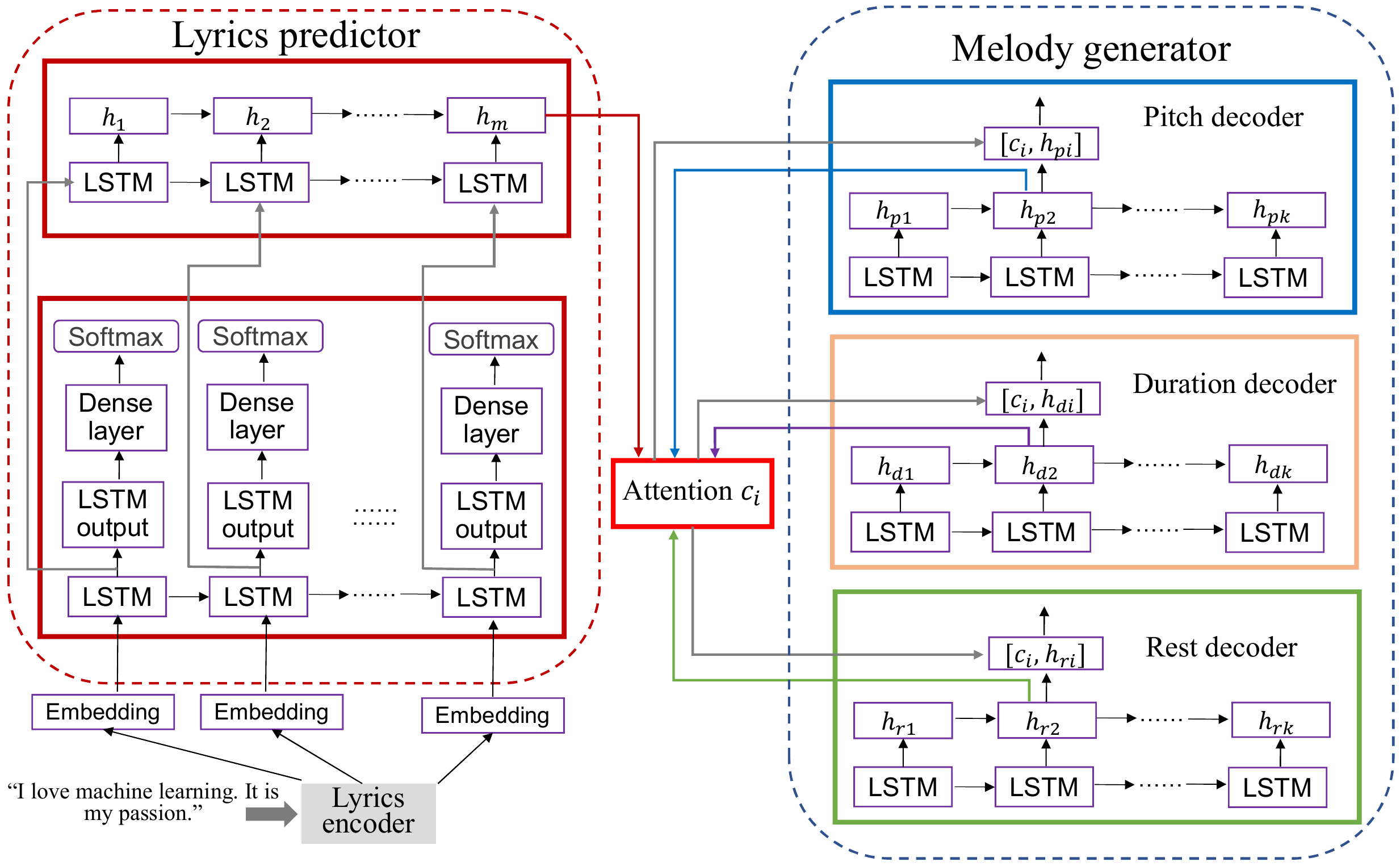}}  
        \caption{Figure shows the proposed Neural lyrics and melody composer (AutoNLMC) model. AutoNLMC consists of a lyrics prediction module, encoder and melody decoders. During inference, lyrics are predicted from the lyrics prediction LSTM module shown in the rectangular red box in the lower half part of the lyrics predictor with the help of softmax layers. While melody is predicted by bypassing the softmax layers to predict pitch, duration and rest with individual decoders as shown in figure.}
        \label{fig:neural_lyrics_melody}
 \end{figure*} 

\section{Lyric to Vector (lyric2vec)}
\label{sec:feat_repre} 

We train continuous skip-gram models to obtain the dense representation of the input lyrics text. This enables us to learn high-quality real valued vector representation of text tokens in an efficient way from large corpus of text data~\cite{mikolov2013efficient}. The vector representations computed from the skip-gram model explicitly encode many linguistic regularities and patterns which can be represented as linear translations. We exploit the linear translation property of embeddings to obtain the lyrics embedding representation to train deep generative models.  

The objective of the skip-gram model is to predict the surrounding context tokens given a token at position $t$ in a piece of input text. We can define the objective function more formally given a sequence of tokens such as $s_1, s_2, . . ., s_T$. We want the model to maximize the log probability~\cite{mikolov2013distributed} of the surrounding tokens given by

\begin{equation}
\frac{1}{T} \sum_{t=1}^{T} \sum_{-c \le i \le c, i \neq 0} \log p({s_{t+i}} | {s_{t}})
\end{equation}
      
\noindent where $c$ is the length of the context of $t^{th}$ token $s_t$. 

The vanilla skip-gram model formulates the objective function as softmax function given by 

\begin{equation}
p(s_O | s_I) = \frac{\exp(v_{s_0}^T  v_{s_I})}{\sum_{t=1}^{T} \exp(v_{s_0}^T  v_{s_I}) } 
\end{equation}

\noindent where  $v_{s_I}$ and  $v_{s_0}$ are input and output vector representations of token $s$, and $T$ is the number of unique tokens in the corpus. This formulation is impractical given the large number of unique tokens/vocabulary of the dataset. Hence, we use negative sampling~\cite{mikolov2013efficient} defined by 

\begin{equation}
\log \sigma(v_{s_O}^T  v_{s_I}) + \sum_{j=1}^{j=k} \E_{s_j \sim p_n(s)} [\log \sigma (v_{s_j}^T  v_{s_I}) ]
\end{equation}  

\noindent Here the objective of the negative sampling function is to discriminate the actual token $s_O$ from the $k$ negative tokens drawn from the smoothed noise distribution~\cite{caselles2018word2vec} $p_n(s)$ given by 

\begin{equation}
p_n(s) = \frac{f(s)^\alpha}{\sum_c f(s)^\alpha}
\end{equation}  
  
\noindent where $f(s)$ is token frequency $s$, and $\alpha$ is real valued distribution smoothing parameter. We train skip-gram models to obtain the embedding vectors of lyrics at syllable and word level. We divide the lyrics of each song into sentences, each sentence into words and each word into syllables. We treat syllables $S = \{s_1, s_2, ..., s_n\}$ as tokens to train the syllable level embedding model and words $W = \{w_1, w_2, w_3, ..., w_n\}$ as tokens for training word level embedding model.
We train skip-gram model as a logistic regression with stochastic gradient decent as optimizer, learning rate  with an initial value 0.03 is gradually decayed every epoch until 0.0007. We use $c = 7$ tokens context window and negative sampling distribution parameter $\alpha$ is set to 0.75. We trained the models to obtain the syllable and world level embedding vectors of dimensions $v = 50$. The skip-gram model to extract the lyric embeddings is shown in Fig.~\ref{fig:skip_gram}. The input one-hot vector representing the syllable or word is compressed to a low dimensional dense representation by linear units to predict the context tokens. 
The following embedding representations are explored individually to train the lyrics prediction and melody composition model: 1) syllable embeddings (SE) where each syllable $s_i$ is encoded with $v$ dimensional vector from syllable embedding model. 2) syllable and corresponding word embedding concatenation (SWC) where a syllable $s_i$ is concatenated with corresponding word embedding $w_i$ of $v$ dimension. 3) addition of syllable and word vector (ASW) where syllable vector is added element wise to the word vector. 4) concatenated syllable, word and syllable projected word vector (CSWP). In CSWP, we project the syllable embedding vector $s_i$ onto the corresponding word vector $w_i$ to give the wordness to syllable $s_i$ by 

\begin{equation}
proj_{w_i} s_i = \frac{s_i.w_i}{|w_i|^2} w_i
\end{equation}         

\noindent finally, we concatenate $s_i$, $w_i$ and $proj_{w_i} s_i$ to form an embedding vector for syllable $s_i$. 



\section{Lyrics and Melody Composer}
\label{sec:prop_lyrics_generator_melody_composer}

The proposed lyrics and melody composer is a sequential encoder-decoder model trained end-to-end to predict lyrics and compose the melody for the predicted lyrics. Initially, AutoNLMC  takes a piece of seed lyrics as input and starts generating the lyrics one sentence at a time with syllables as tokens. For each generated lyrics, the melody is composed until the last sentence. At the end, all pairs of generated lyrics and melody are stitched together to form the complete song. 


  
\subsection{Neural Lyrics Generator and Melody Composer}
\label{subsec:neural_lyric_melody}

The proposed neural lyrics and melody composer (AutoNLMC) is shown in Fig.~\ref{fig:neural_lyrics_melody}. The AutoNLMC is a sequential RNN encoder-decoder network consists of a lyric generator, encoder and melody decoder. The RNN sequential lyrics generator and encoder are clubbed together during training as shown in Fig.~\ref{fig:neural_lyrics_melody}. The melody decoder is designed to decode the melody attributes in the form of MIDI. Specifically, the lyric generator takes the sequence of lyric tokens $S$ as input where each token is embedded into a fixed size dense vector by the skip-gram model discussed in section~\ref{sec:feat_repre} to generate the next lyrics token. Each generated lyrics token is encoded by a RNN encoder to a fixed dimensional hidden vector $h_t$ at each time step. The melody decoder uses the encoded hidden vector $h_t$ and the dynamic context vector $C_t$ to generate the melody decoder hidden state at each time step to generate the melody sequence. The decoder consists of three independent MIDI decoders for each attribute receives the lyric context vector $C_t$ and decoder state to generate the pitch, duration and rest MIDI attributes for each syllable. We design independent decoder for each attribute to reduce the number of classes and hence to increase the accuracy of the predicted output, which is partially motivated by the separate pitch and rhythm models in~\cite{ackerman2017algorithmic}. Each unit in the RNN is modeled with gated LSTM~\cite{cho2014learning}. The hidden state $h_t$  of the lyrics generator is computed by 

\begin{equation}
h_t = (1-z_t) \circ h_{(t-1)} + z_t \circ \bar{h}_t
\end{equation} 

\begin{equation}
\bar{h}_t = \tanh(W s_t-1 + U [r_i \circ h_{t-1}])
\end{equation}

\begin{equation}
z_t = \sigma (W_z s_t + U_z h_{t-1})
\end{equation}             

\begin{equation}
r_t = \sigma(W_r s_t + U_r h_{t-1})
\end{equation}

\noindent where $z_t$ and $r_t$ are the reset and update gates of the LSTM unit. $\sigma$ is the non-linear sigmoid function. The reset gate makes the hidden state to forgot the past sequence information irrelevant to predict the future sequence. While, the update gate controls the information flow from previous hidden states to current state, and further acts as memory cell to remember long term sequential dependencies which helps to learn the semantic information of sequences. The variables $W, W_z, W_r, U, U_z, U_r$ are the weights of the LSTM unit. The weights are learned automatically during training. The lyrics generator is trained to maximize the log conditional probability 

\begin{equation}
\operatorname*{max}_\theta \frac{1}{N} \sum_{i=1}^{N} \log p_\theta(s_i | s_{< i})
\end{equation}

\noindent where $s_i$ denotes the target lyrics token of the $i^{th}$ example and  $s_{< i}$ is shorthand notation for $\{s_1, s_2, ..., s_{t-1}\}$.   

The melody decoder decodes the melody attributes $m_i = \{p_i, d_i, r_i\}$ i.e., pitch, duration, and rest independently by maximizing the log conditional probability 

\begin{equation}
\operatorname*{max}_\theta \frac{1}{N} \sum_{i=1}^{N} \log p_\theta(m_i | s_i)
\end{equation}

\noindent The decoder states $\tilde{h}_t$ where $\tilde{h}_t = \{\tilde{h}_{pt}, \tilde{h}_{dt}, \tilde{h}_{rt} \}$ are the states of individual decoders given the encoder hidden state $h_t$ is computed by

\begin{equation}
\tilde{h}{_t} = (1-z_t) \circ \tilde{h}_{t-1} + z_t \circ \bar{\tilde{h}}_t
\end{equation} 

\begin{equation}
\bar{\tilde{h}}_t = \tanh(W m_{t-1} + U [r_i \circ \tilde{h}_{t-1}] + C c_t)
\end{equation}

\begin{equation}
z_t = \sigma (W_z m_{t-1} + U_z \tilde{h}_{t-1} + C_z c_t)
\end{equation}             

\begin{equation}
r_t = \sigma(W_r m_{t-1} + U_r \tilde{h}_{t-1} + C_r c_t)
\end{equation}

\noindent The free variables $W, W_z, W_r, U, U_z, U_r, C, C_z, C_r$ are all learnable parameters. 
 
The attention vector $c_t$ to align the input lyrics with the melody by computing alignment model~\cite{bahdanau2014neural}

\begin{equation}
c_t = \sum_{j=1}^{T} \alpha^{tj} h_t 
\end{equation} 

\begin{equation}
\alpha^{tj} = \frac{exp(e^{tj})}{\sum_{k = 1}^{T} exp(e_{tk})}
\end{equation} 

\begin{equation}
e^{tj} = V^{T}_a \tanh(W_a \tilde{h}_{t-1} + U_a h_t)
\end{equation}

\noindent the parameters $V, W_a, U_a$ are the learnable weights.

The neural lyric and melody composition model is trained end-to-end to minimize the total loss function defined by
 
\begin{equation}
\begin{split}
L = - \frac{1}{N} \sum_{i=1}^{N} \sum_{j=1}^{|M^{(i)}|} \log p (M^{(i)} | \theta, S^{(i)}, m_{<k}^{(i)}, c^{(i)}) \\ 
 -\frac{1}{N} \sum_{i=1}^{N} \sum_{k=1}^{|S^{(i)}|} \log p(S^{(i)k} | S^{(i)}_{<k})
\end{split}
\end{equation}
 
\begin{equation}
\begin{split}
 L = - \frac{1}{N} \sum_{i=1}^{N} \sum_{k=1}^{|M^{(i)}|} \log p (p^{(i)k}, d^{(i)k}, r^{(i)k} | \theta, S^{(i)}, m_{<k}^{(i)}, c^{(i)}) \\  -\frac{1}{N} \sum_{i=1}^{N} \sum_{k=1}^{|S^{(i)}|} \log p(S^{(i)k} | S^{(i)}_{<k})
 \end{split}
\end{equation}

\begin{equation}
\begin{split}
L = - \frac{1}{N} \sum_{i=1}^{N} \bigg( \sum_{j=1}^{|M^{(i)}|} \log p (p^{(i)k} | \theta, S^{(i)}, p_{<k}^{(i)}, c^{(i)}) \\ + \sum_{j=1}^{|M^{(i)}|} \log p (d^{(i)k} | \theta, S^{(i)}, d_{<k}^{(i)}, c^{(i)}) \\ + \sum_{j=1}^{|M^{(i)}|} \log p (r^{(i)k} | \theta, S^{(i)}, r_{<k}^{(i)}, c^{(i)}) \bigg) \\  -\frac{1}{N} \sum_{i=1}^{N} \sum_{k=1}^{|S^{(i)}|} \log p(S^{(i)k} | S^{(i)}_{<k})
\end{split}
\end{equation}

\noindent where $N$ is the set of  $(S^{(i)}, M^{(i)})$ lyric melody pairs, $S^{(i)} = \{s^{(i)1}, s^{(i)2}, ...., s^{|S^{(i)}|}\}$, $M^{(i)} = \{m^{(i)1}, m^{(i)2}, ...., m^{|M^{(i)}|}\}$ and  $m^{(i)1} = \{p^{(i)1}, d^{(i)1}, r^{(i)1} \}$.

The number of LSTM units used for all the models is 128. 
The initial state of lyrics generator model is initialized with zero vector, where as the encoder initial state is initialized with the last hidden state of lyrics generator model. The pitch decoder, duration decoder and rest decoder of the melody prediction model is initialized with the last hidden state of encoder model. The loss function is minimized by Adam optimizer with initial learning rate of 0.0001 and linearly decayed after every 10 epochs. The model is trained to minimize the cross entropy loss function with 32 batch size. All weight matrices are initialized from zero mean, 0.02 variance Gaussian distribution.

\subsection{Lyrics and Melody Inference}
\label{subsec:inference}

Lyrics inference is a process of generating new lyrics by combining the previously generated lyrics from the conditional probability distribution to generate the whole lyrics. 
Specifically, we obtain the next lyric token by multinomial distribution output by the softmax non-linear activation function as shown in the lower part of lyrics predictor of Fig.~\ref{fig:neural_lyrics_melody}. At each time step, we generate a lyrics token and feedback to the RNN unit at next time step to generate new token. We use the following strategy to generate the lyrics from the lyrics prediction model and evaluate the effectiveness of each strategies in the evaluation section. i) Greedy search: at each time step, we pick the most probable token from the softmax probability distribution given by 

\begin{equation}
\operatorname*{arg\,max}_j \ p(s_{(t,j)} = 1 | s_{(t-1)}, ..., s_{1}) = \frac{\exp(w_j h_t)}{\sum_{j^{'}=1}^{K} exp(w_{j^{'}} h_t)} 
\end{equation}

\noindent where $j = 1,..., K$ are the unique lyric tokens, $w_j$ are the rows of the LSTM weight matrix $W$. ii) Temperature sampling: The probability of each lyric token is transformed to a freezing function $f_\tau$~\cite{Russell} with a controllable parameter $\tau $ before sampling most likely lyric token given by 

\begin{equation}
p_j = f_\tau(p_j) = \frac{p_j^{\frac{1}{\tau}}}{\sum_{j=1}^{K} p_j^{\frac{1}{\tau}}}  
\end{equation}   

\noindent the value of $\tau \in [0, 1]$ makes the model to predict more robust and diverse lyrics.

The melody inference to predict the most probable melody sequence of the generated lyrics is done at two modes: i) composing melody from the lyrics generated from the lyrics generator and ii) composing melody from the original lyrics created by the human lyrics writer. In the former case, we obtain the lyrics from the lyrics generator. For each lyric token, we obtain the dense representation from the skip-gram model. The attention vector at each time step which aligns the input lyrics to the output pitch is computed from the encoder hidden vectors. The initial state of the pitch decoder is initialized with the last state of the encoder. At each time step, we feed the variable attention vector and the decoder hidden state to the softmax probability distribution function to predict the pitch corresponding to the lyric token. To generate the full length sequence at each time step, we feed the previously generated pitch tokens to generate the next tokens. We iterate this process until we generate all pitch tokens corresponds to lyrics. Similarly, we follow the same steps to decode the duration and rest attributes of all lyric tokens. Finally, we club the predicted melody triplets i.e., pitch, duration and rest and assign to the corresponding lyric syllable to form the lyrics melody pairs. In case of composing melody from the original lyrics created by the human lyrics writer, we feed the melody inference model with the lyrics created by the human writer (here, lyrics prediction model is turned off.), where one sentence at a time is generated to match the corresponding melody form the model.            

\begin{figure}[htbp]
        \centering
		\resizebox{8cm}{5cm}{\includegraphics{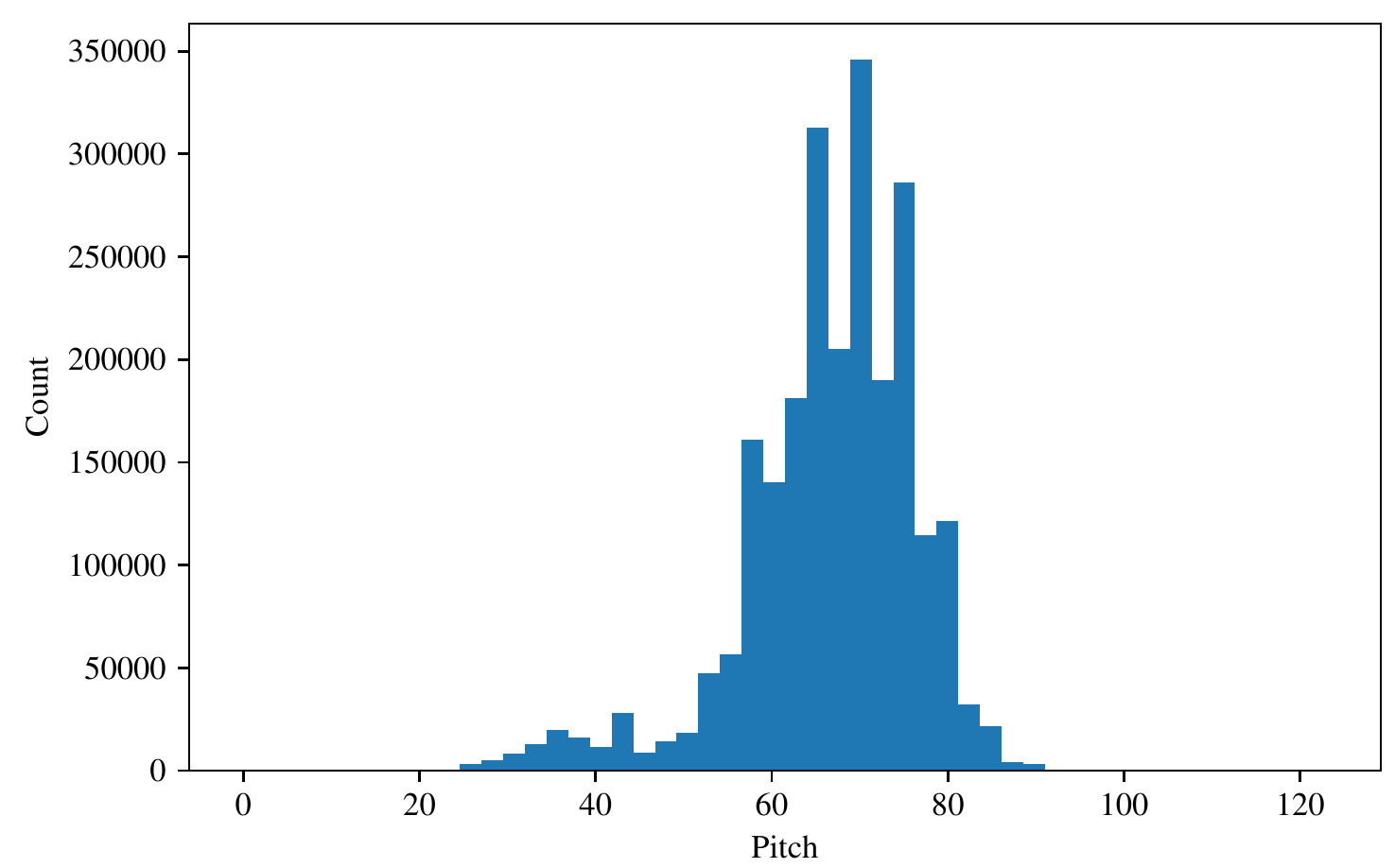}}  
        \caption{Pitch distribution.}
        \label{fig:dataset_pitch_dist}
 \end{figure} 

\begin{figure}[htbp]
        \centering
		\resizebox{8cm}{5cm}{\includegraphics{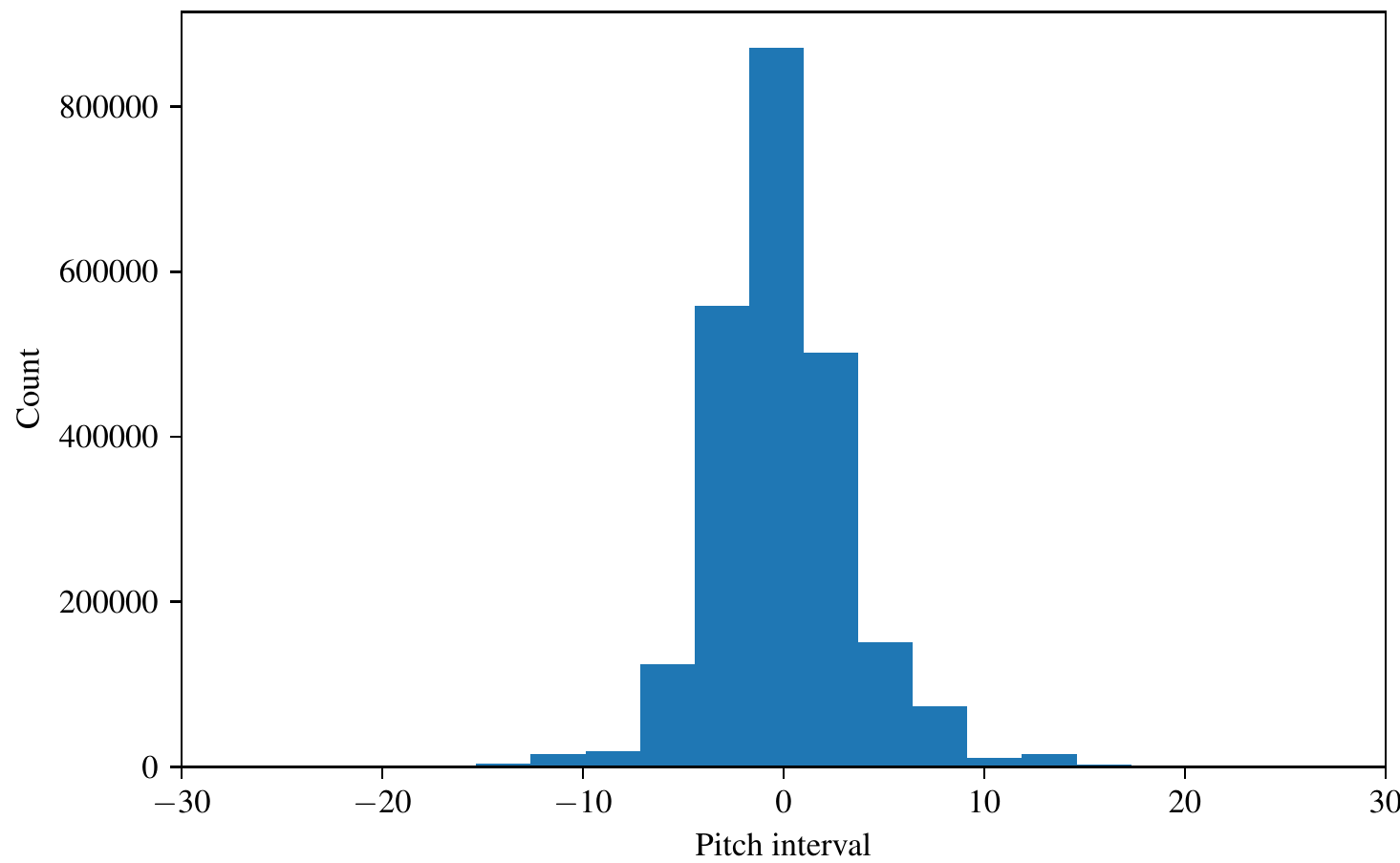}}  
        \caption{Pitch interval distribution.}
        \label{fig:dataset_pitch_interval}
 \end{figure} 

\begin{figure}[htbp]
        \centering
		\resizebox{8cm}{5cm}{\includegraphics{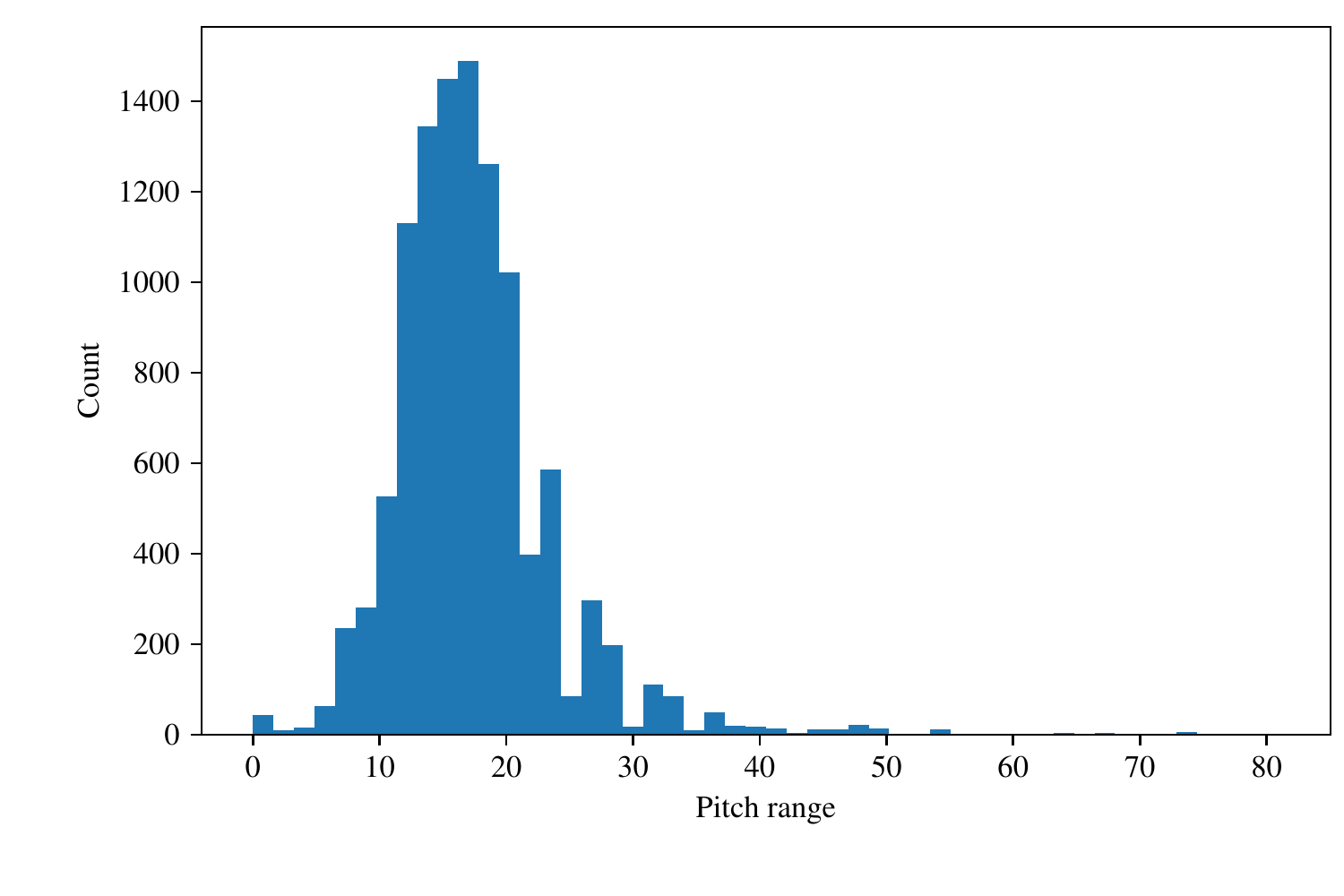}}  
        \caption{Pitch range distribution.}
        \label{fig:dataset_pitch_range}
 \end{figure}

\begin{figure}[htbp]
        \centering
		\resizebox{8cm}{5cm}{\includegraphics{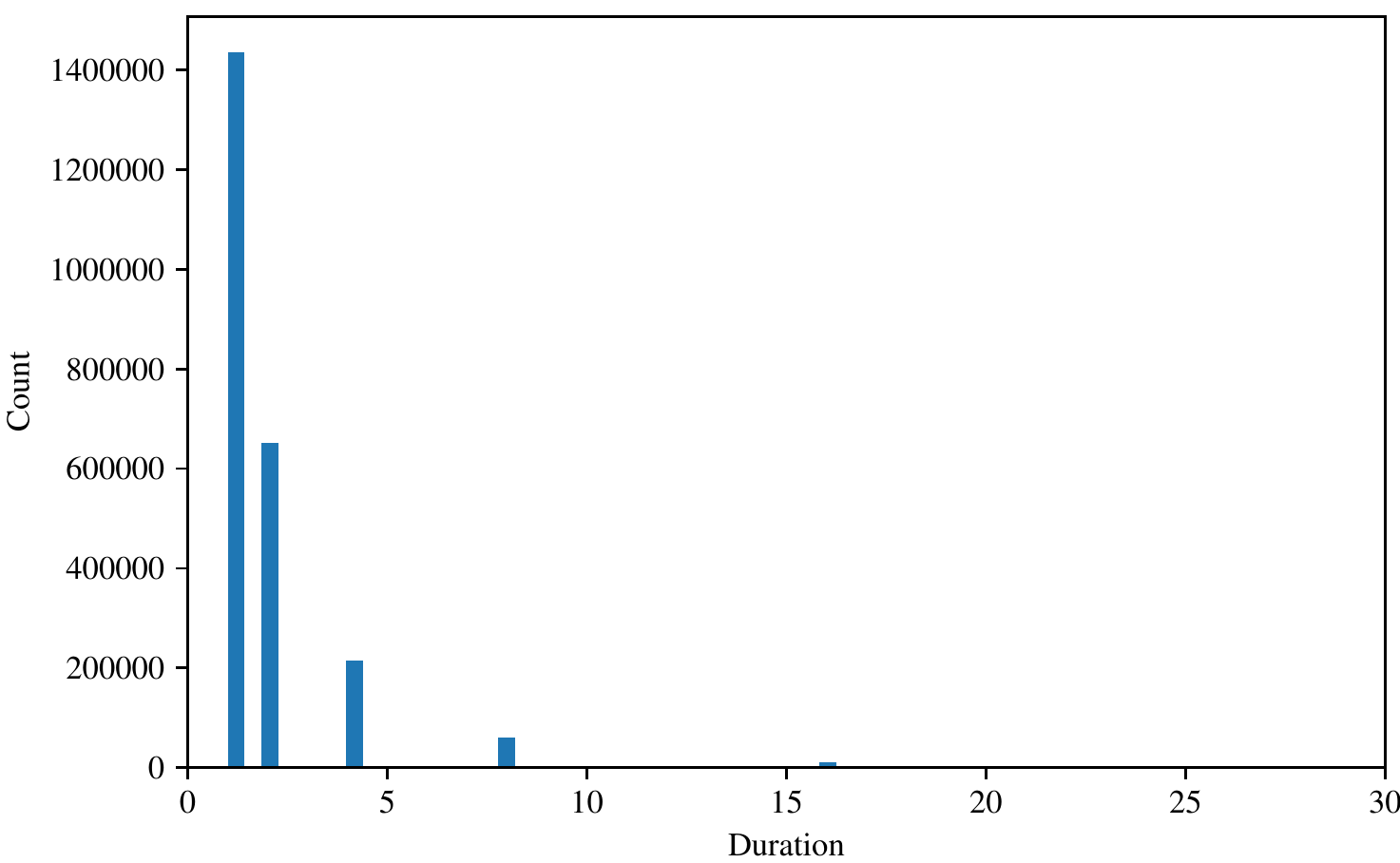}}  
        \caption{Histogram distribution of duration.}
        \label{fig:dataset_duration}
 \end{figure} 

\begin{figure}[htbp]
        \centering
		\resizebox{8cm}{5cm}{\includegraphics{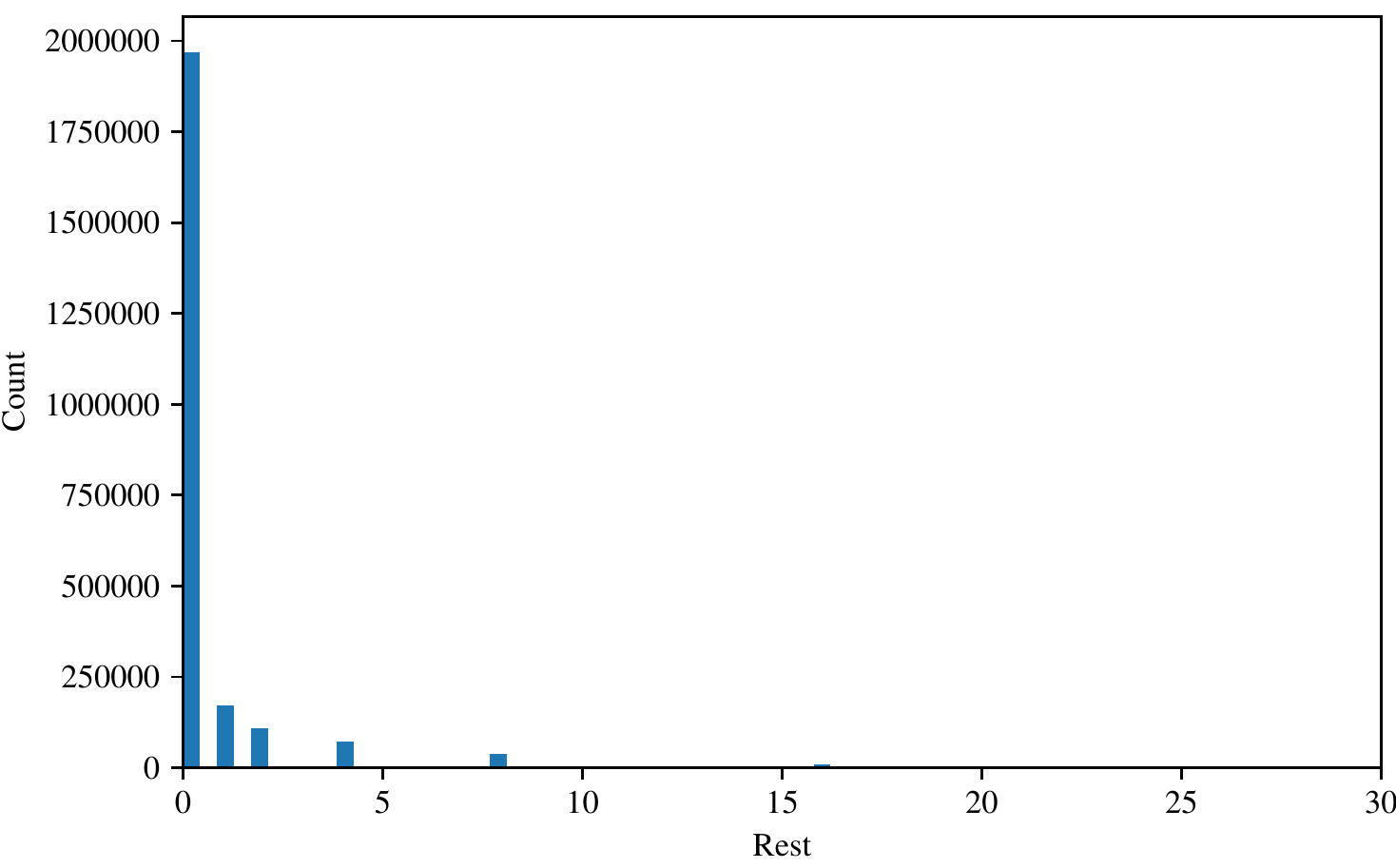}}  
        \caption{Histogram distribution of rest.}
        \label{fig:dataset_rest}
 \end{figure}

\begin{figure*}[htbp]
    \centering
    \subfloat[Greedy search]{{\includegraphics[height=3.5cm, width=4cm]{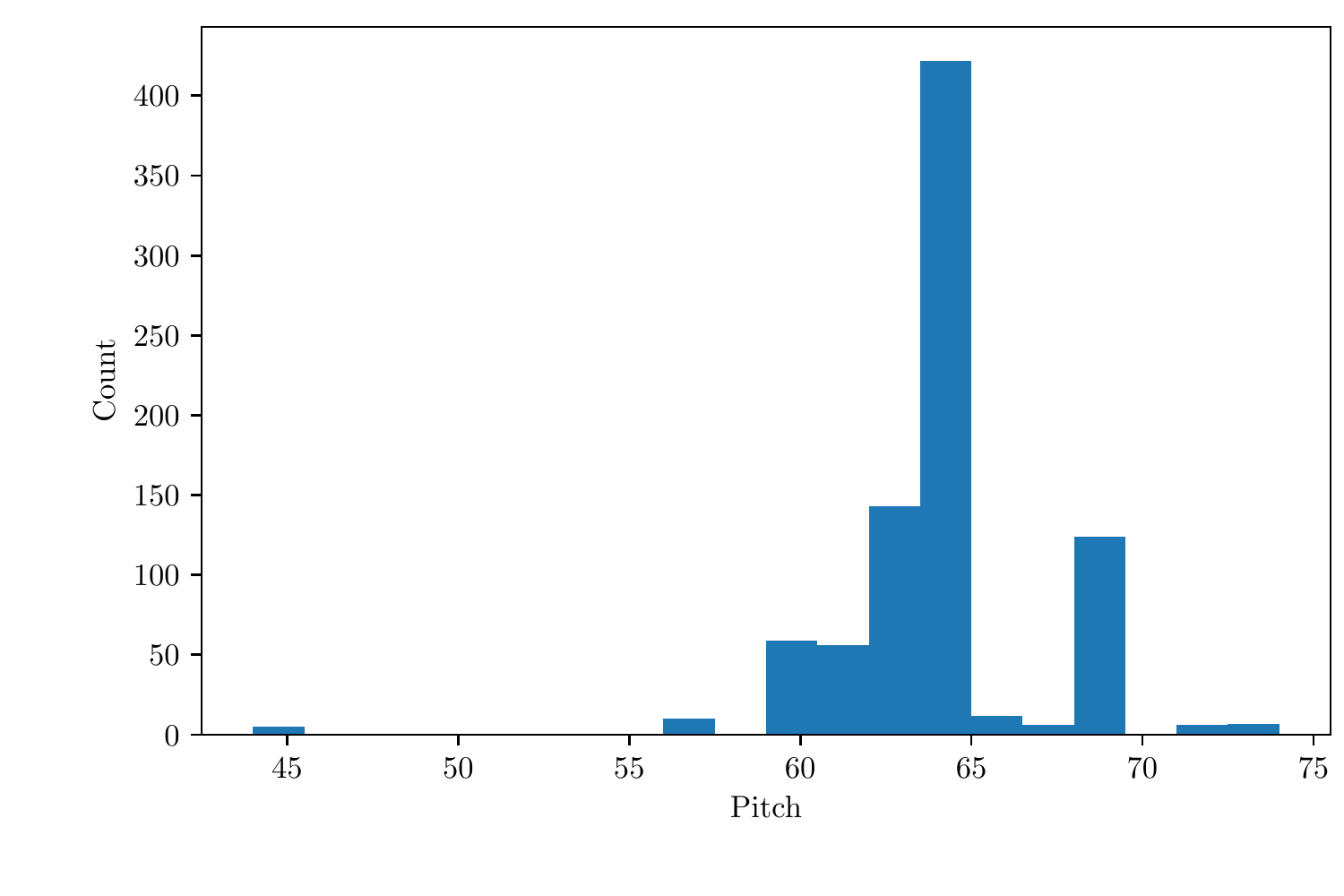}}}%
    \subfloat[$\tau = 0.5$]{{\includegraphics[height= 3.5cm, width=4cm]{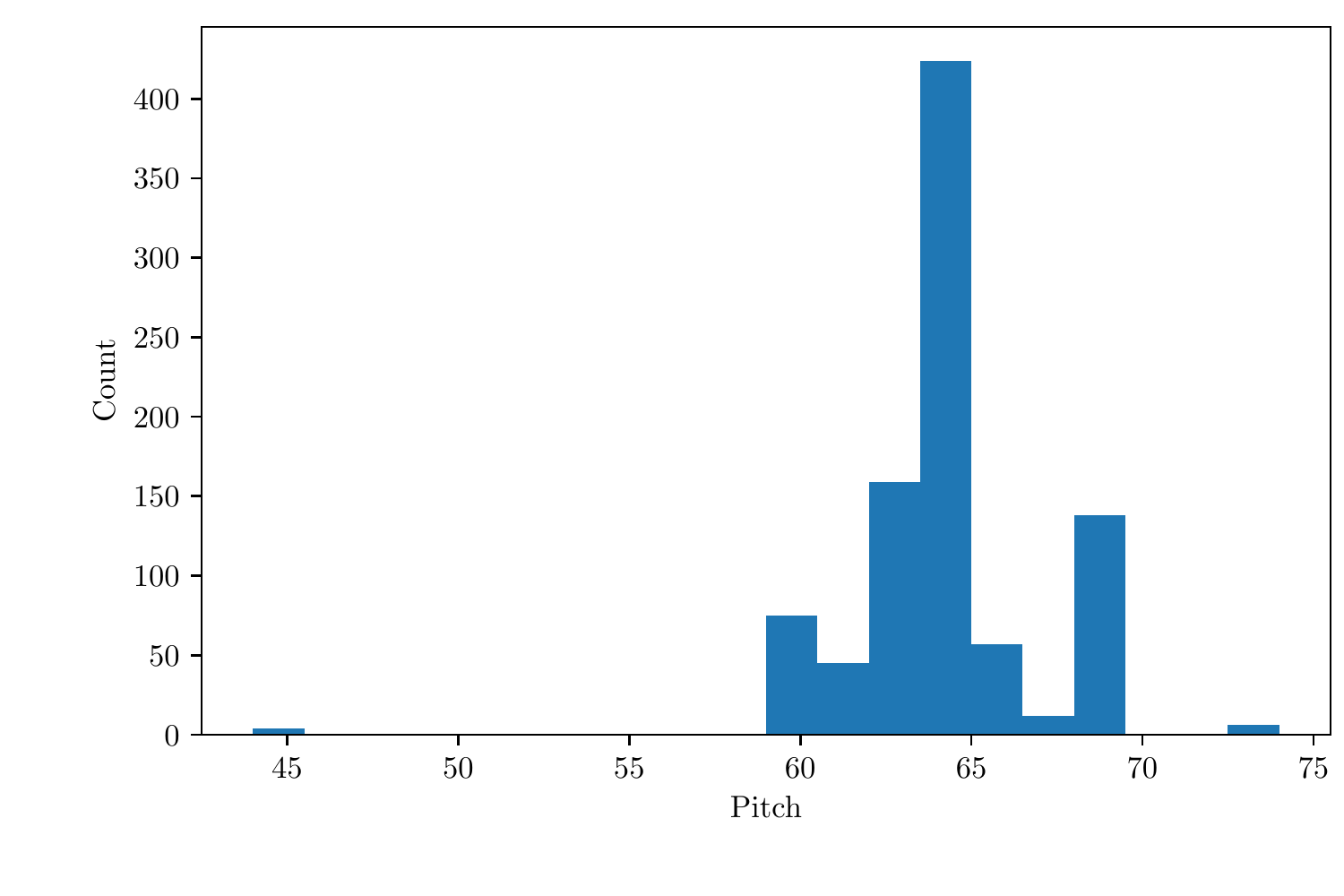}}}%
    \subfloat[$\tau = 0.6$]{{\includegraphics[height= 3.5cm, width=4cm]{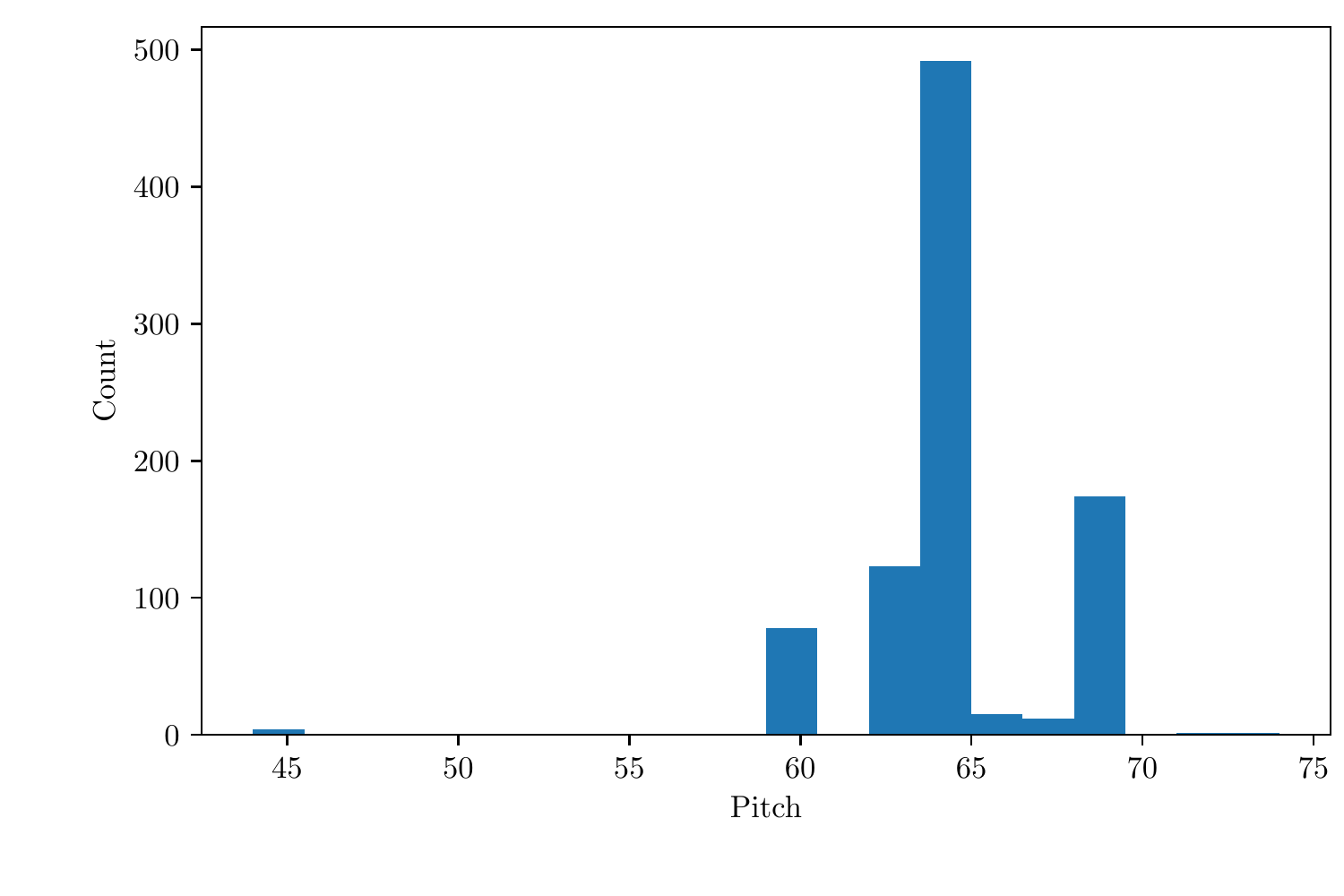}}}%
	\subfloat[$\tau = 0.7$]{{\includegraphics[height= 3.5cm, width=4cm]{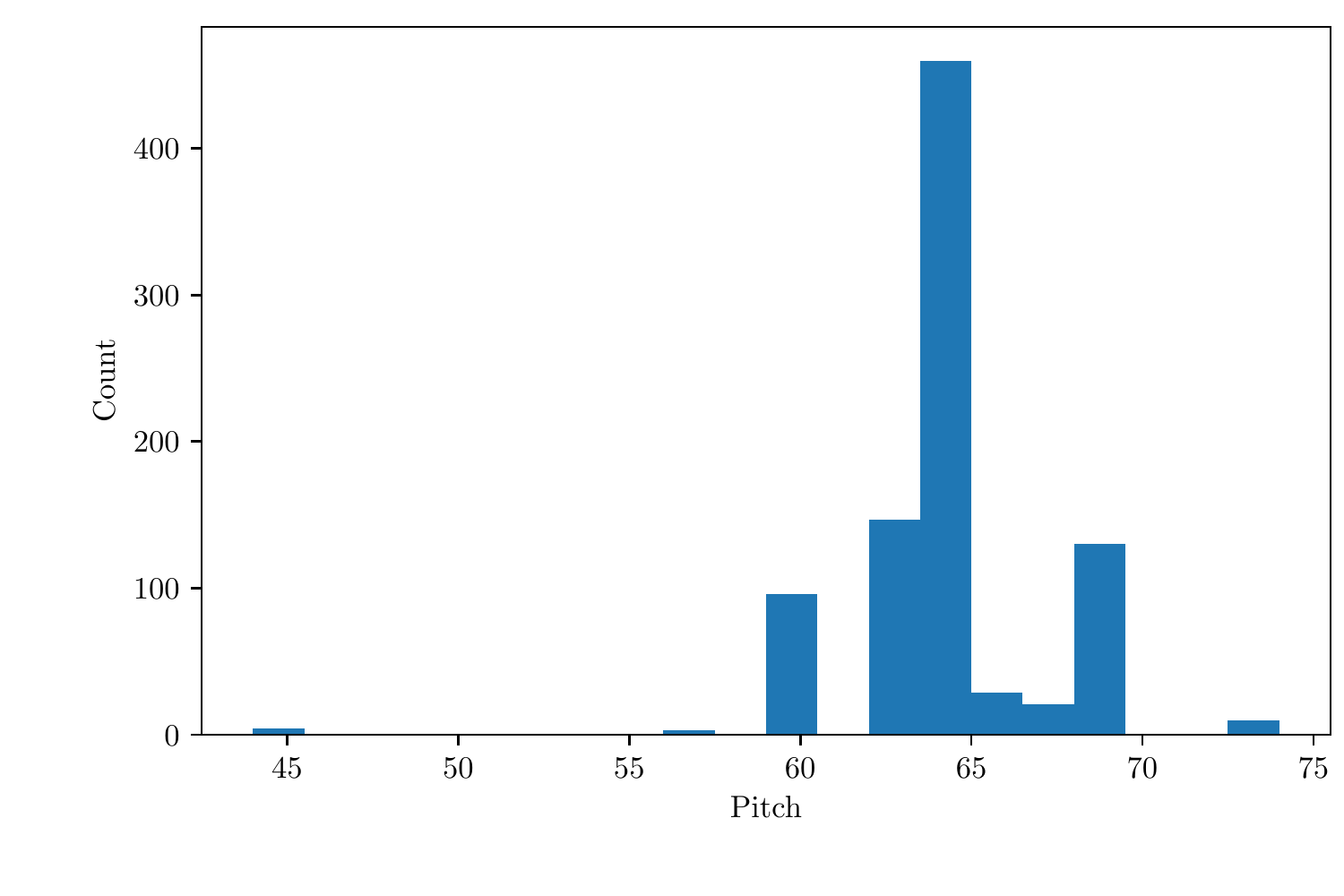}}}\\%
    \subfloat[$\tau = 0.8$]{{\includegraphics[height= 3.5cm, width=4cm]{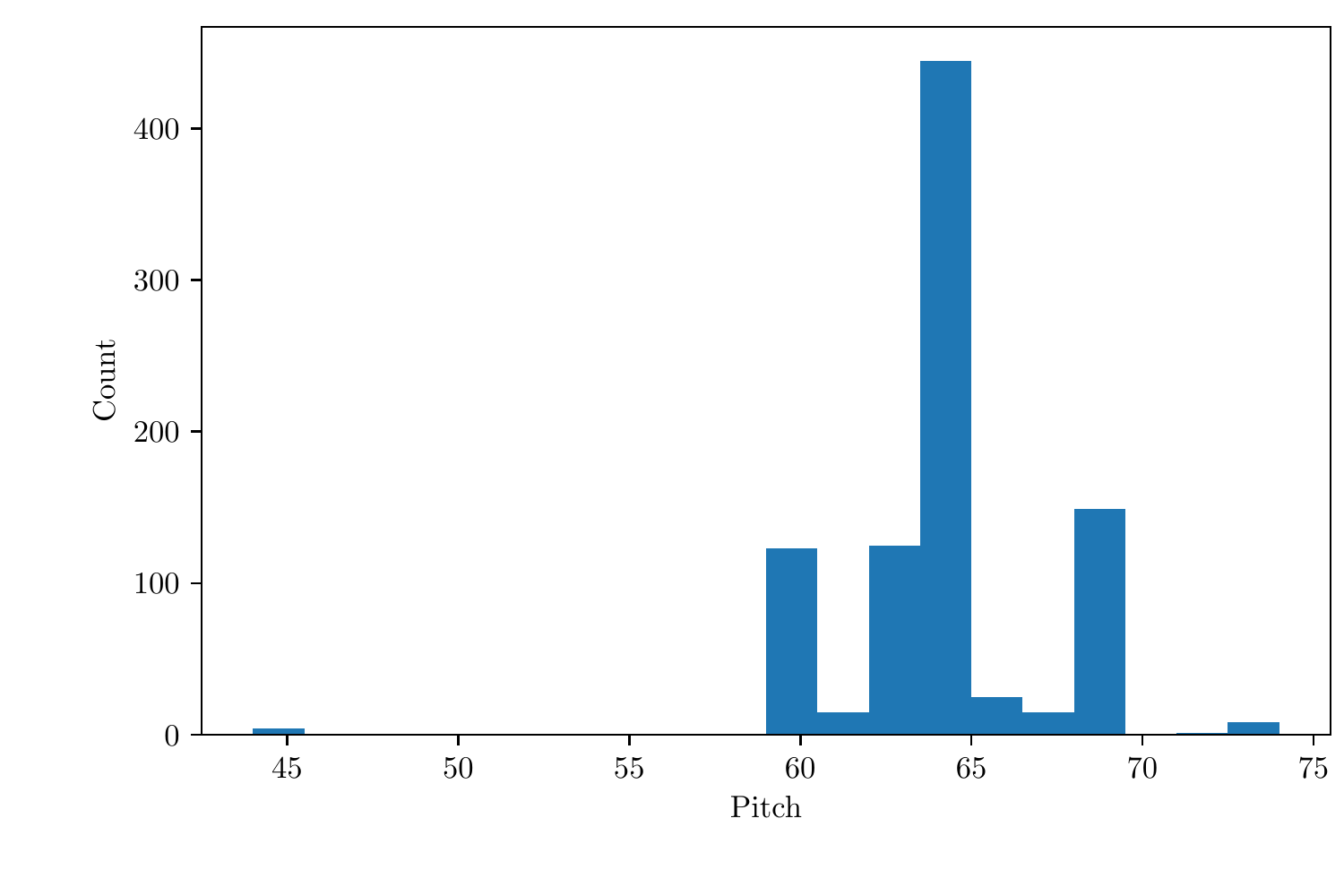}}}%
    \subfloat[$\tau= 0.9$]{{\includegraphics[height= 3.5cm, width=4cm]{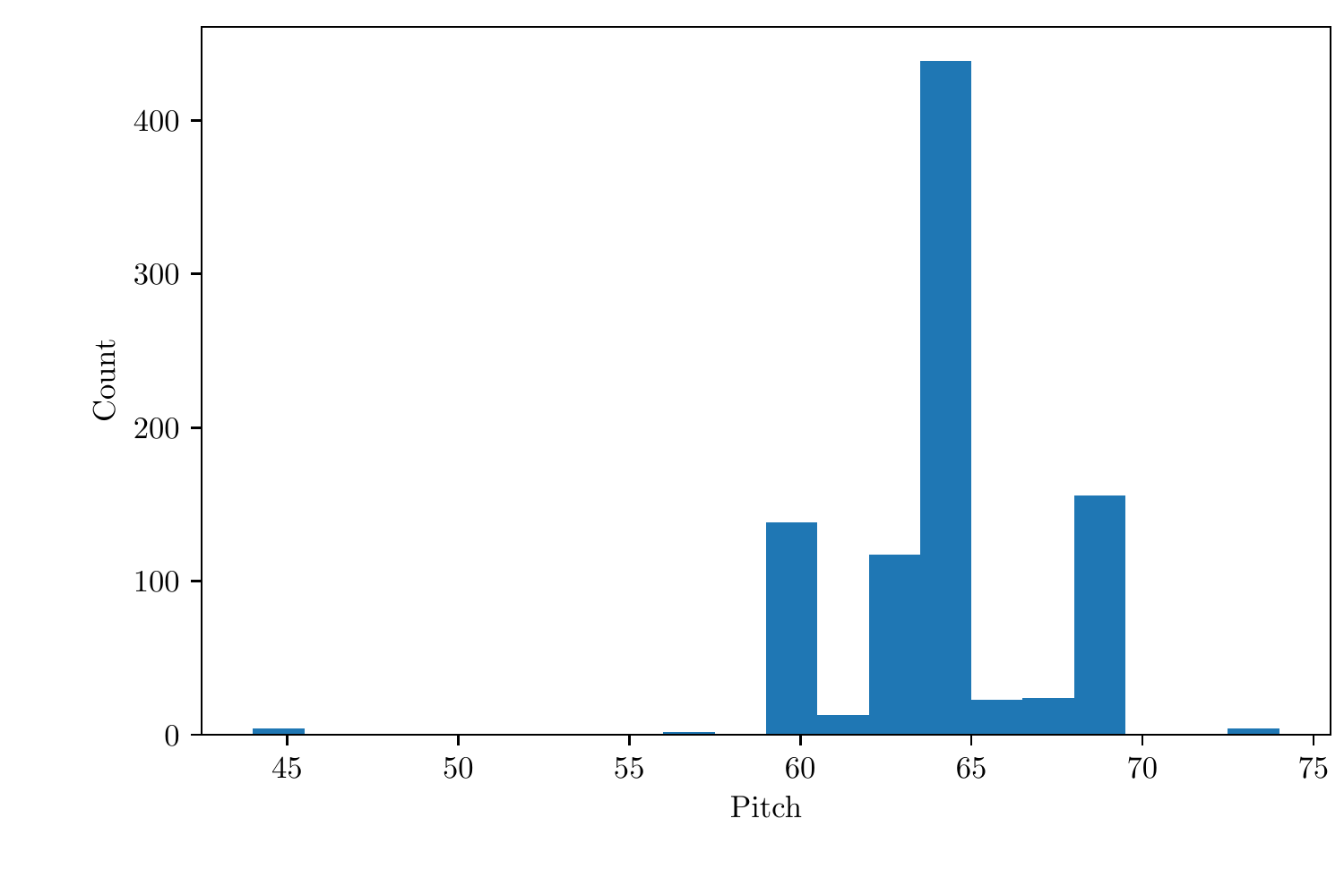}}}%
    \subfloat[$\tau = 1.0$]{{\includegraphics[height= 3.5cm, width=4cm]{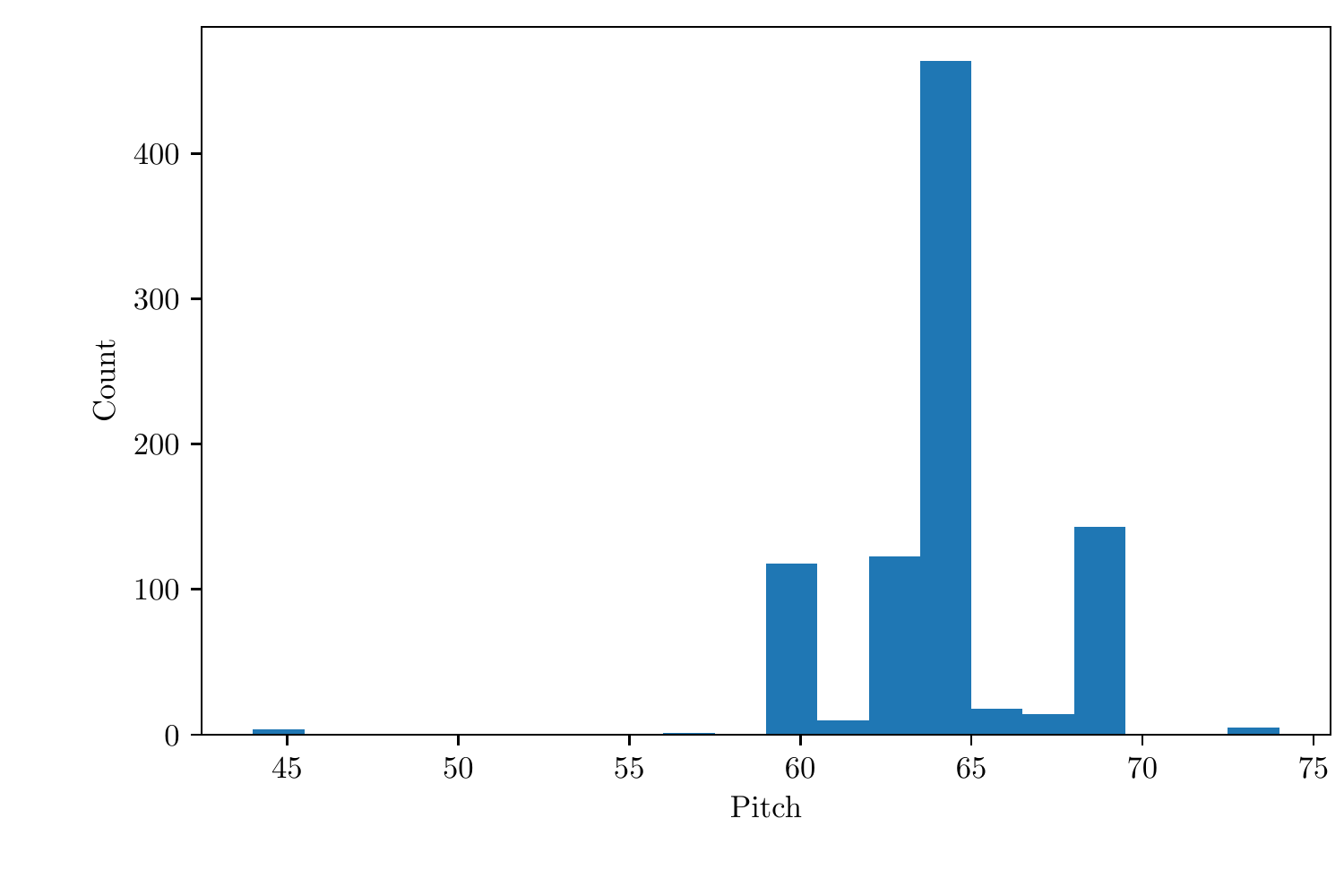}}}%
    \caption{Pitch histogram distribution of the generated lyrics predicted from melody decoder for greedy and temperature sampling methods with lyrics encoded only with syllable embeddings (SE).}%
    \label{fig:generated_pitch_syll_emb}%
\end{figure*}


\begin{figure*}[htbp]
    \centering
    \subfloat[Greedy search]{{\includegraphics[height=3.5cm, width=4cm]{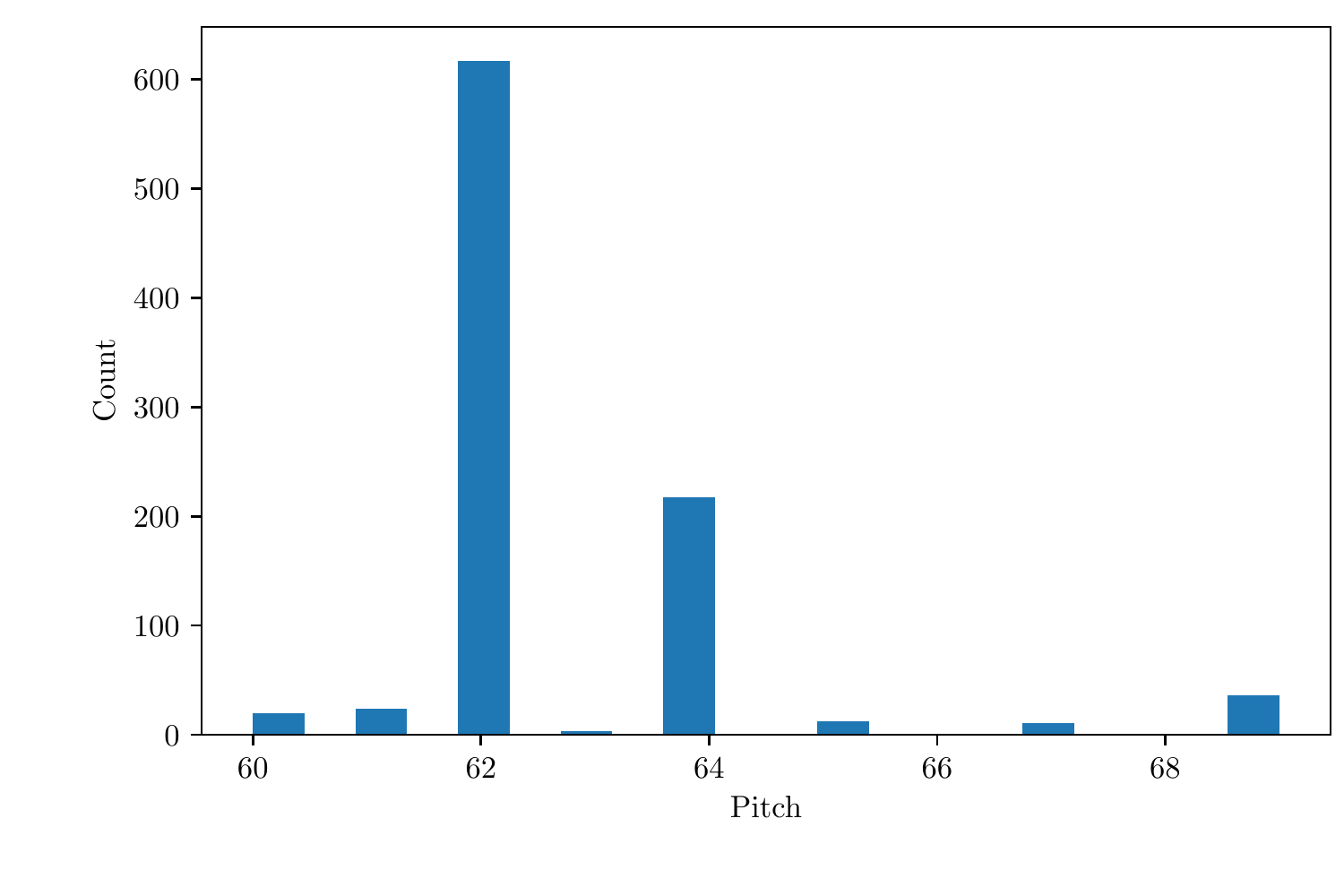}}}%
    \subfloat[$\tau = 0.5$]{{\includegraphics[height= 3.5cm, width=4cm]{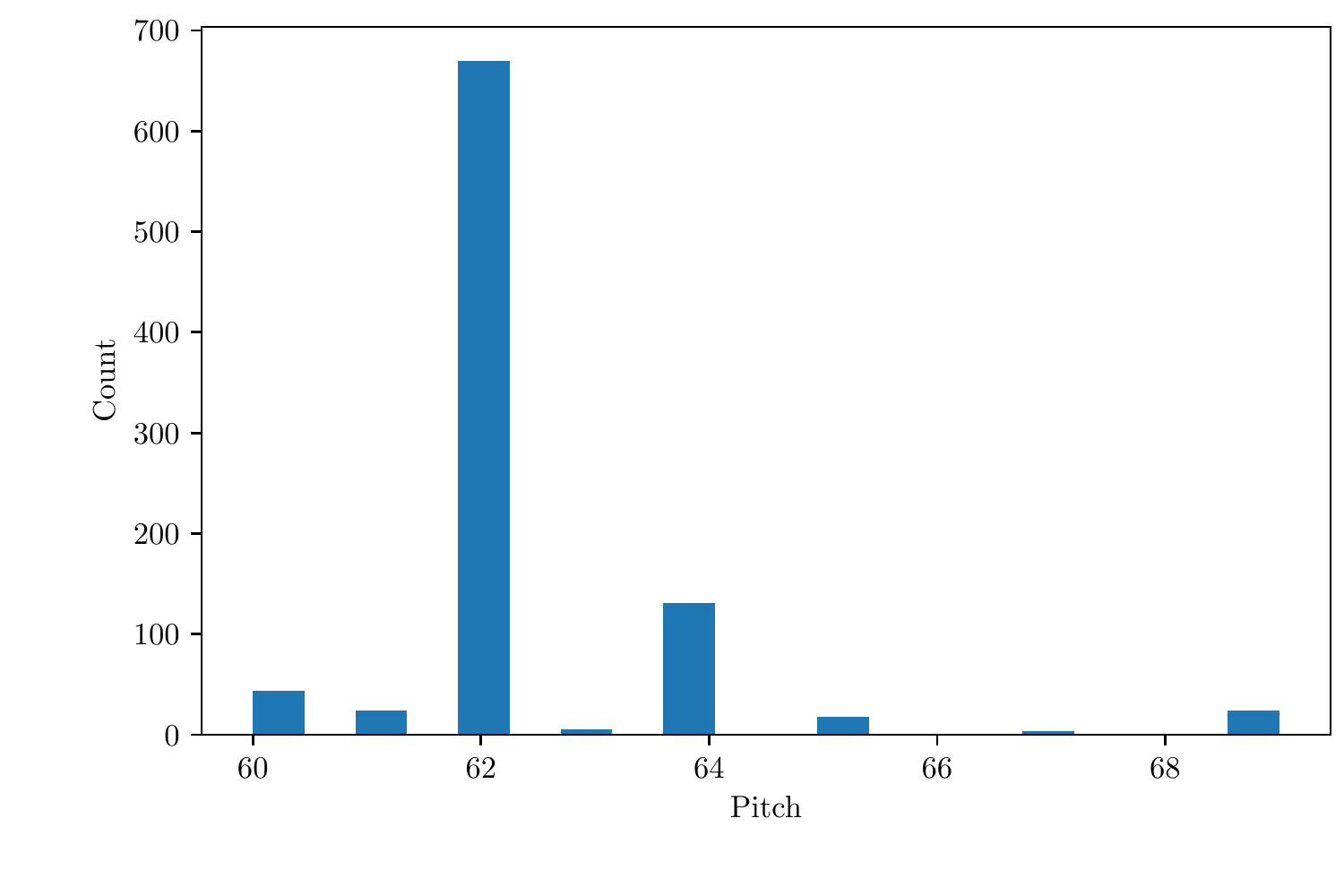}}}%
    \subfloat[$\tau = 0.6$]{{\includegraphics[height= 3.5cm, width=4cm]{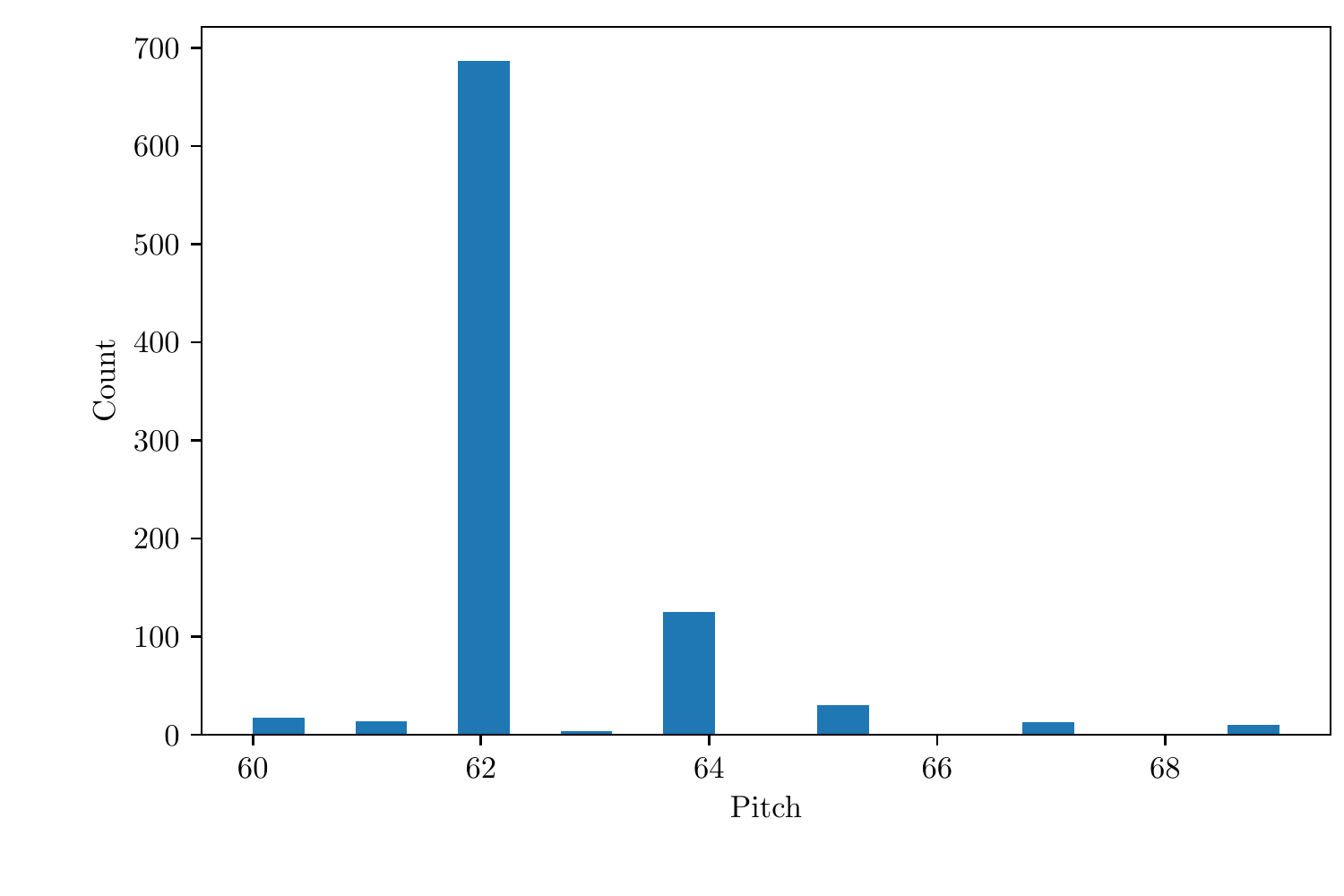}}}%
	\subfloat[$\tau = 0.7$]{{\includegraphics[height= 3.5cm, width=4cm]{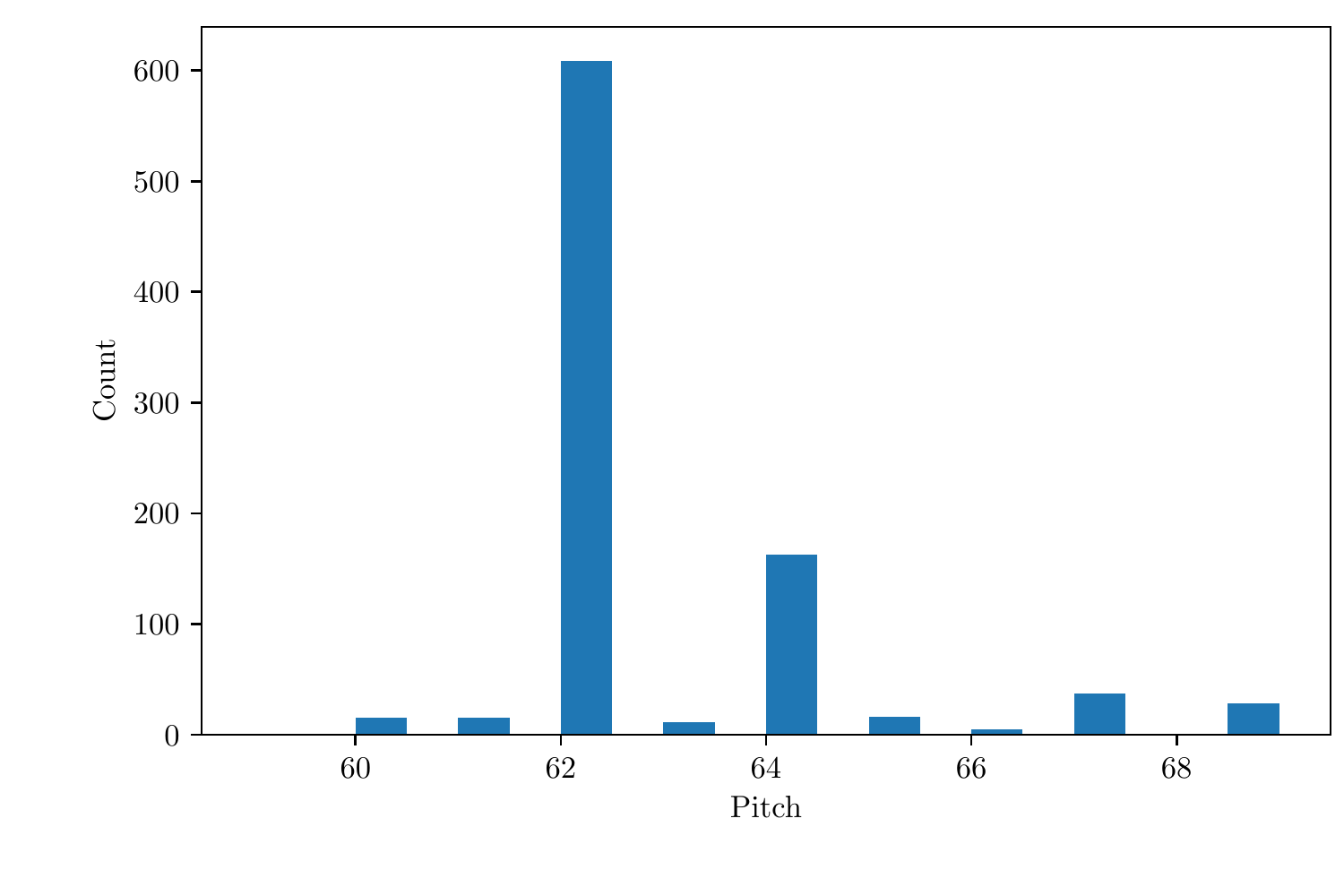}}}\\%
    \subfloat[$\tau = 0.8$]{{\includegraphics[height= 3.5cm, width=4cm]{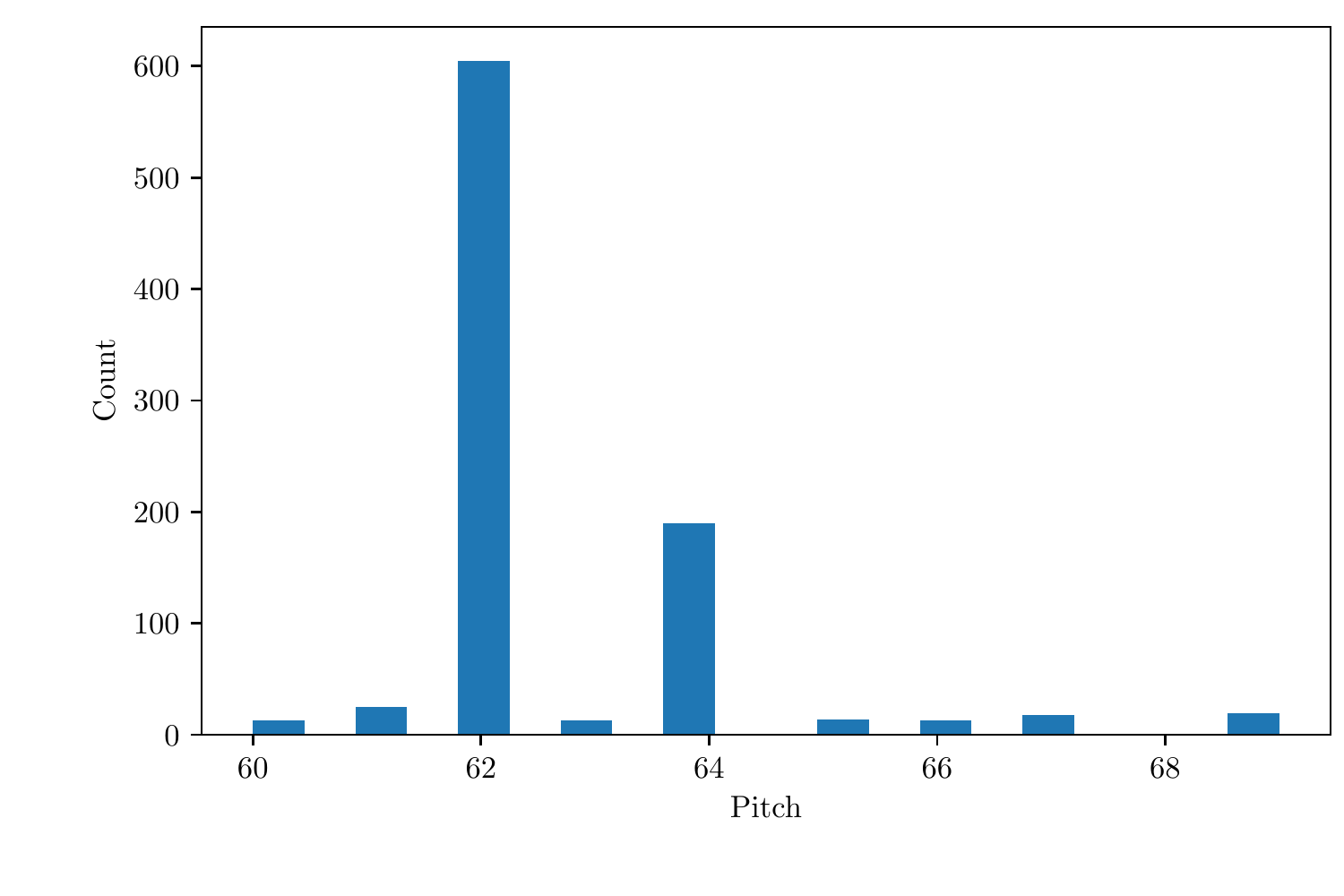}}}%
    \subfloat[$\tau= 0.9$]{{\includegraphics[height= 3.5cm, width=4cm]{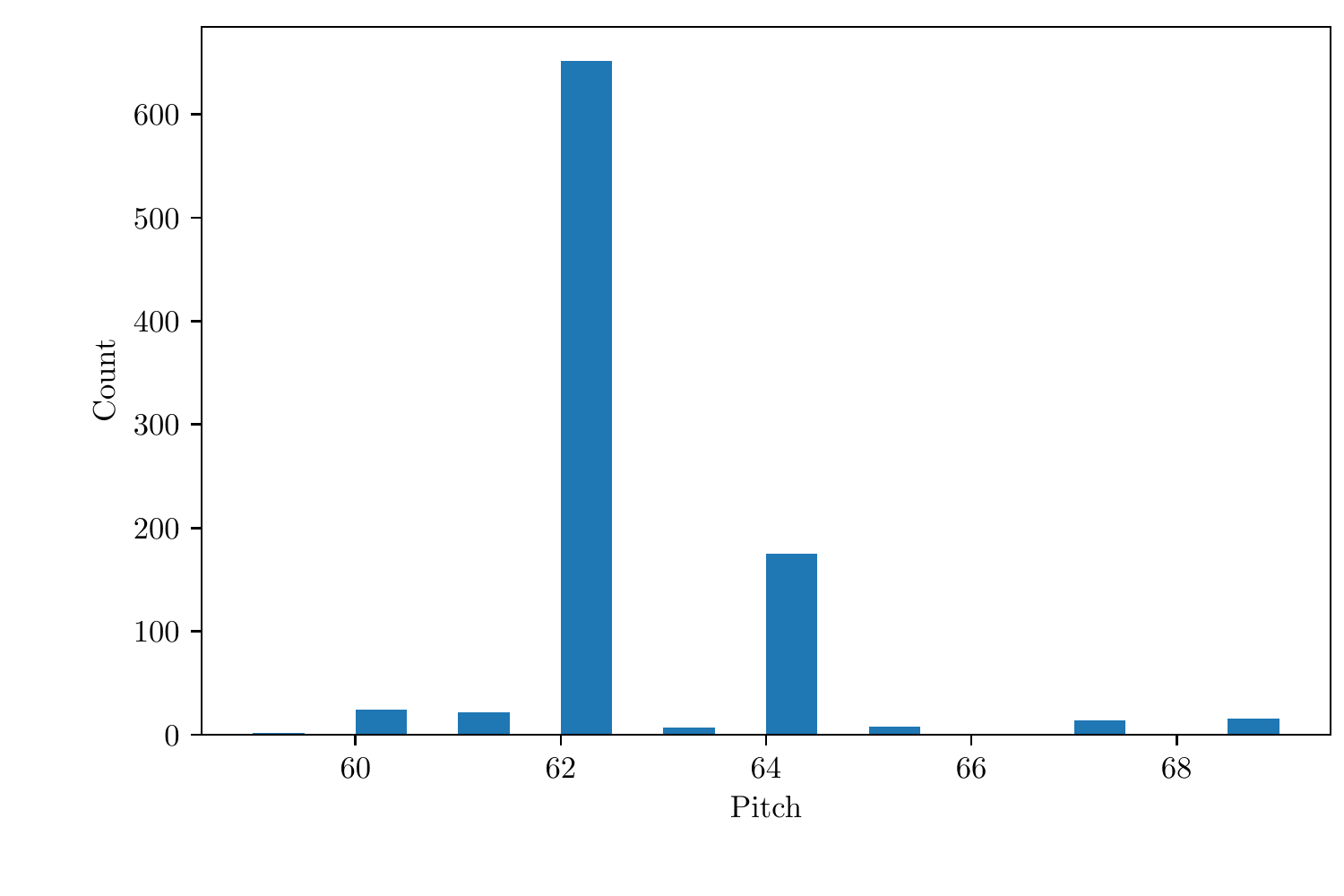}}}%
    \subfloat[$\tau = 1.0$]{{\includegraphics[height= 3.5cm, width=4cm]{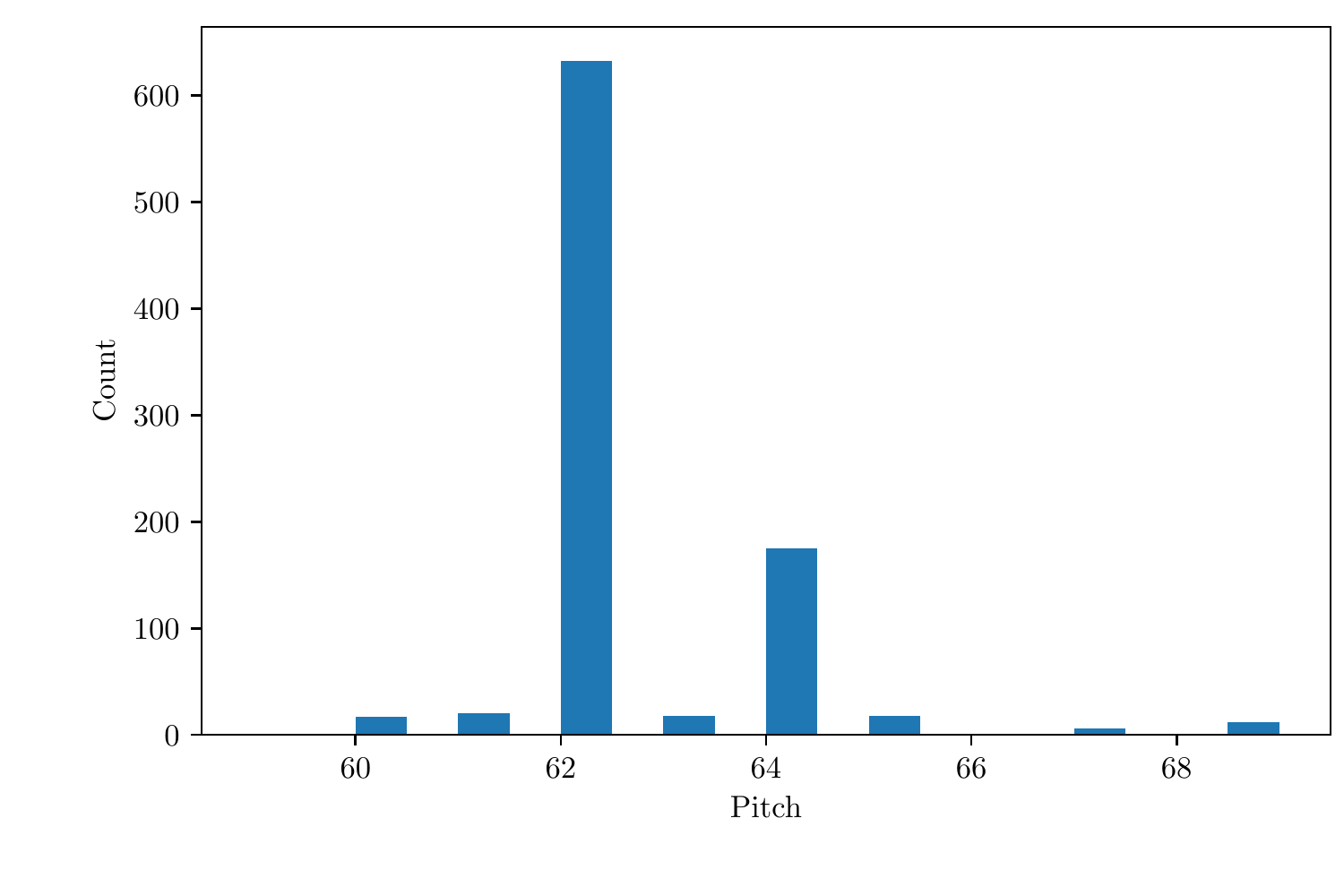}}}%
    \caption{Pitch histogram distribution of the generated lyrics predicted from melody decoder for greedy and temperature sampling methods with lyrics encoded by concatenating syllable and word embeddings (SWC).}%
    \label{fig:generated_pitch_syll_word_concat_emb}%
\end{figure*}


\begin{figure*}[htbp]
    \centering
    \subfloat[Greedy search]{{\includegraphics[height=3.5cm, width=4cm]{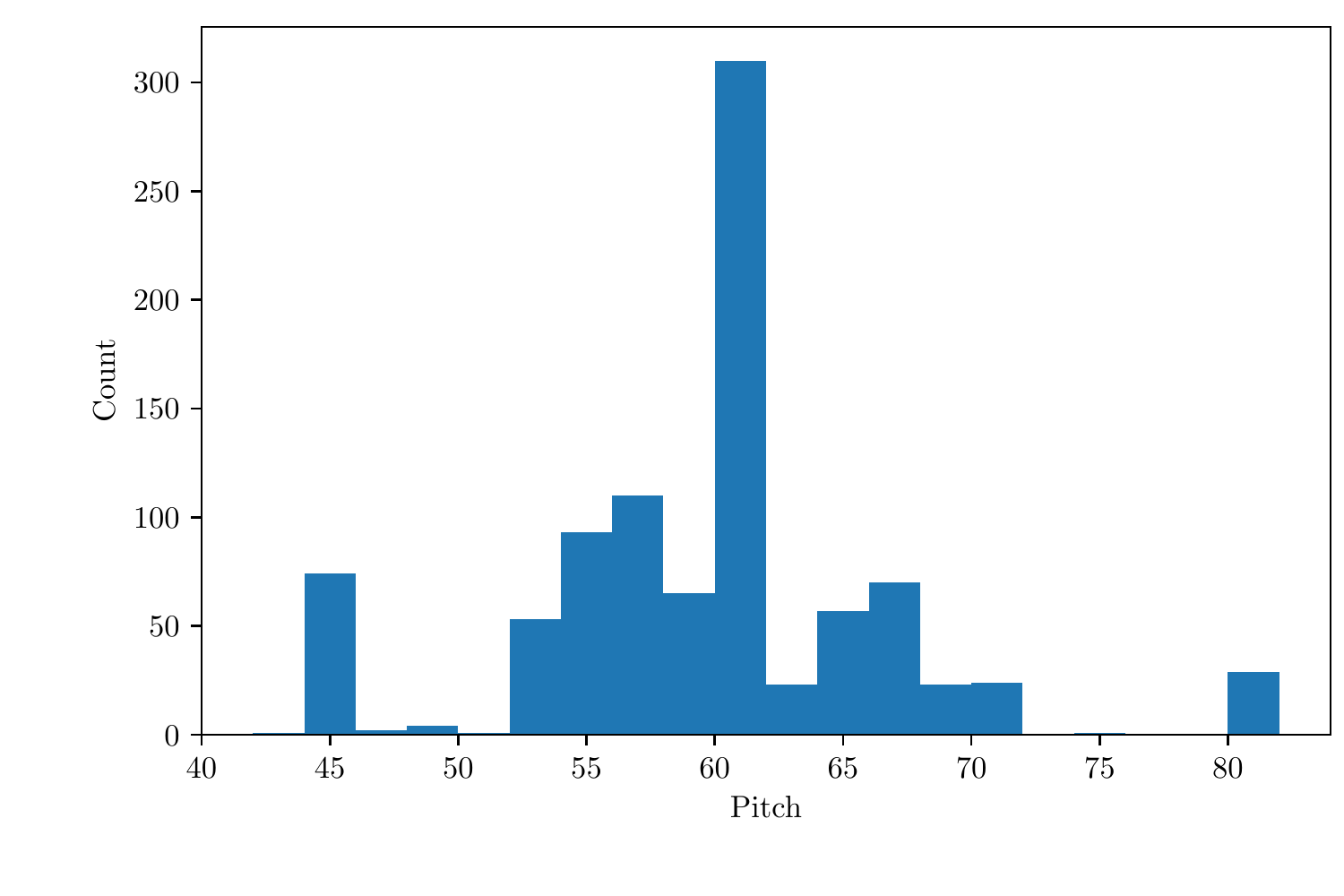}}}%
    \subfloat[$\tau = 0.5$]{{\includegraphics[height= 3.5cm, width=4cm]{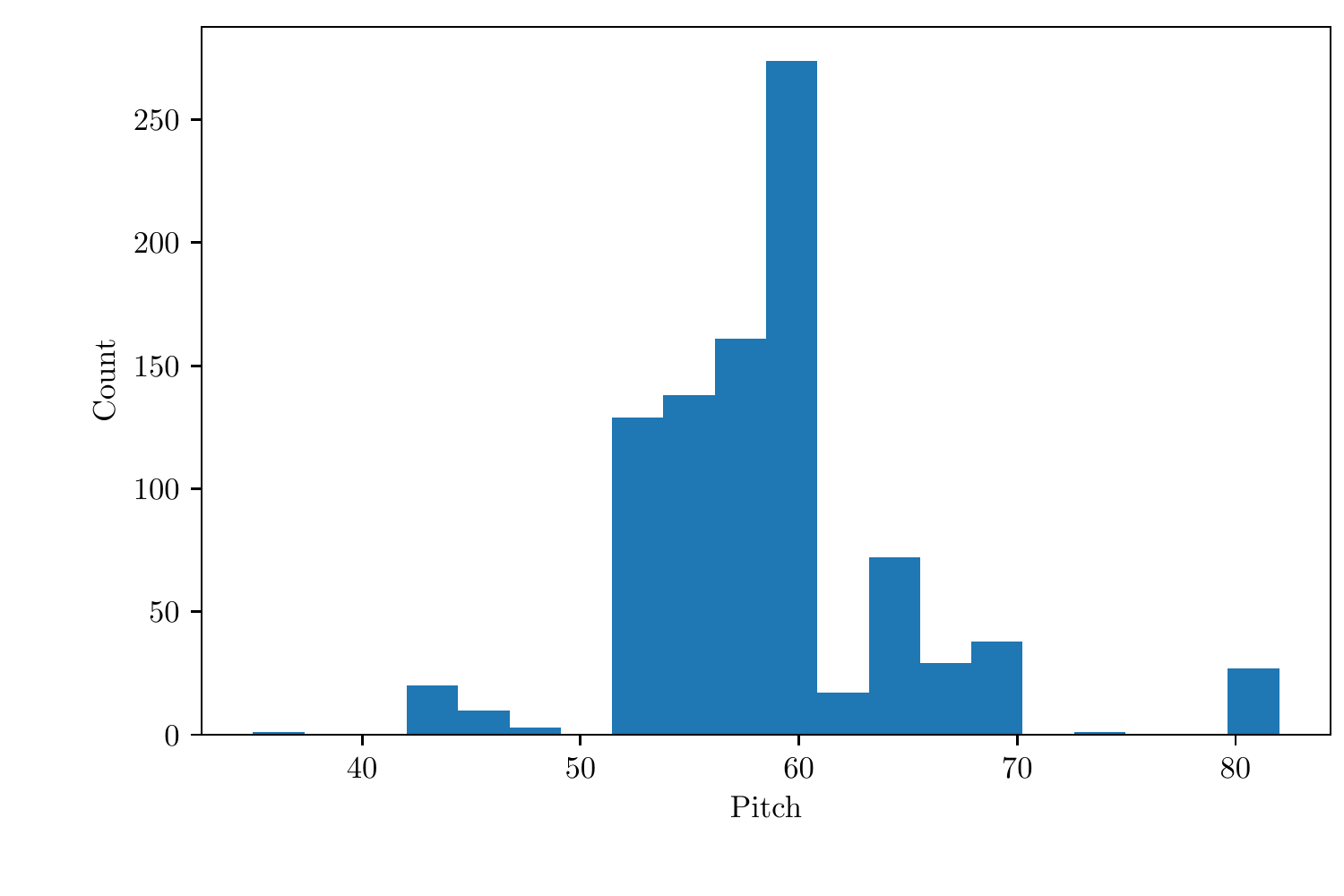}}}%
    \subfloat[$\tau = 0.6$]{{\includegraphics[height= 3.5cm, width=4cm]{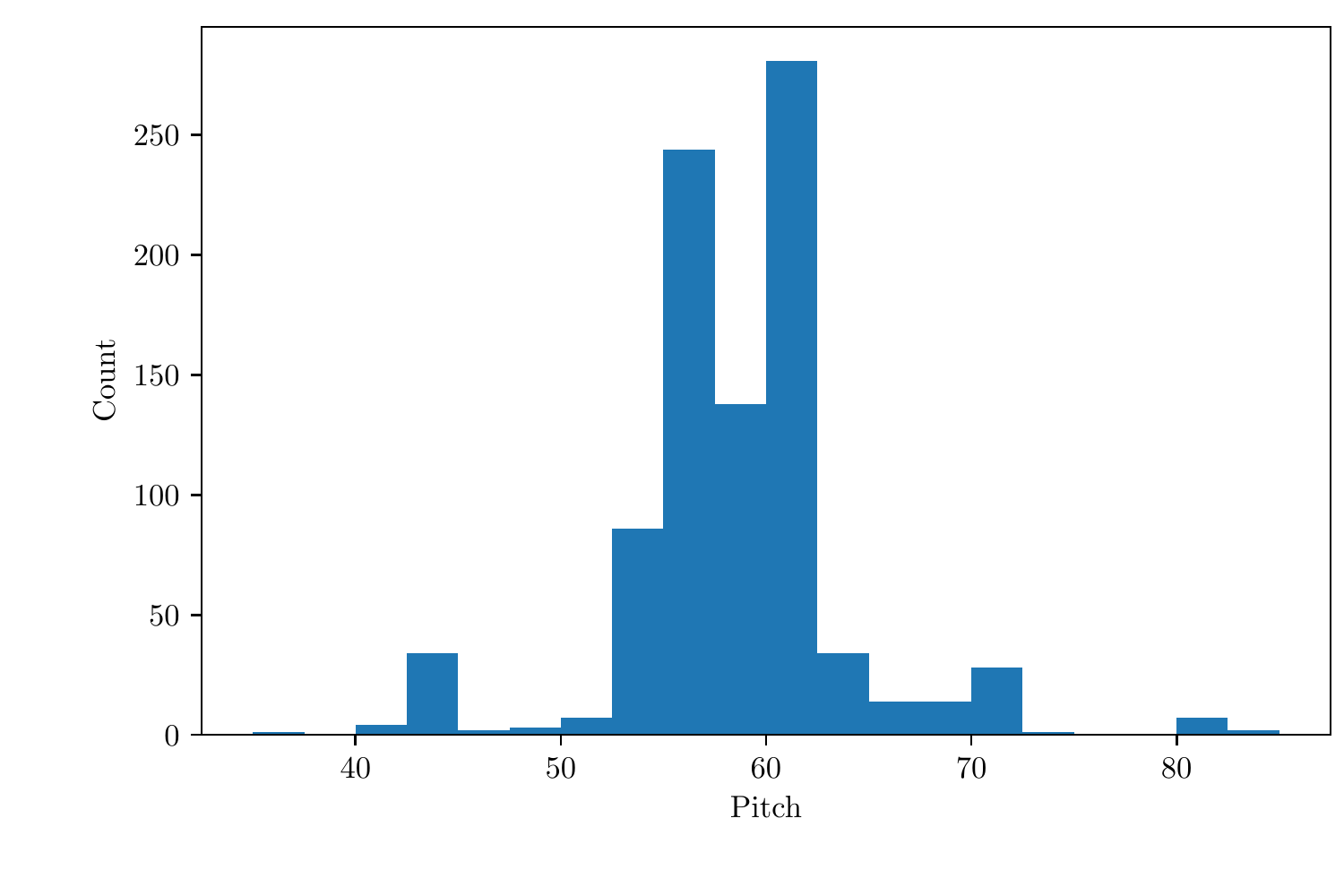}}}%
	\subfloat[$\tau = 0.7$]{{\includegraphics[height= 3.5cm, width=4cm]{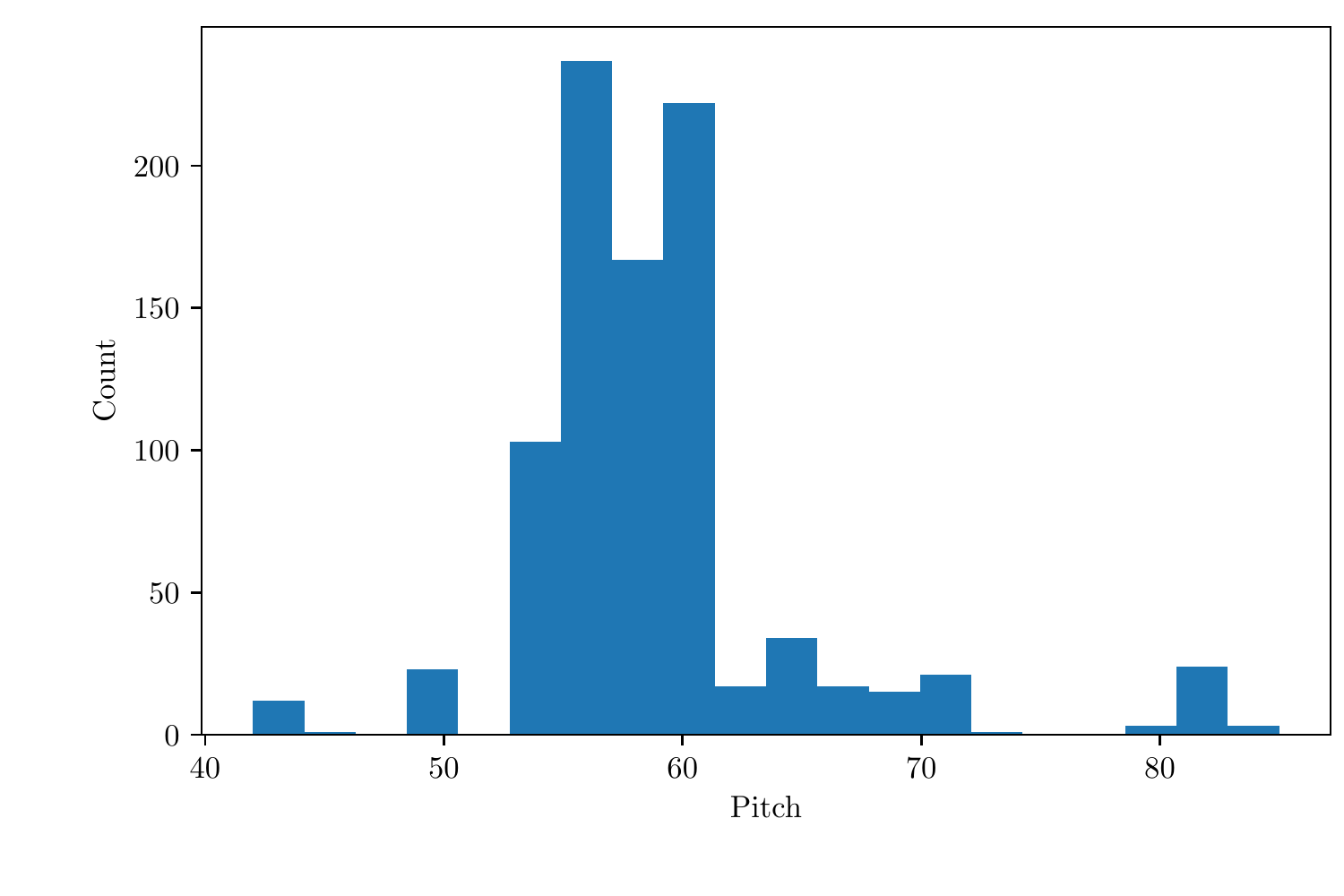}}}\\%
    \subfloat[$\tau = 0.8$]{{\includegraphics[height= 3.5cm, width=4cm]{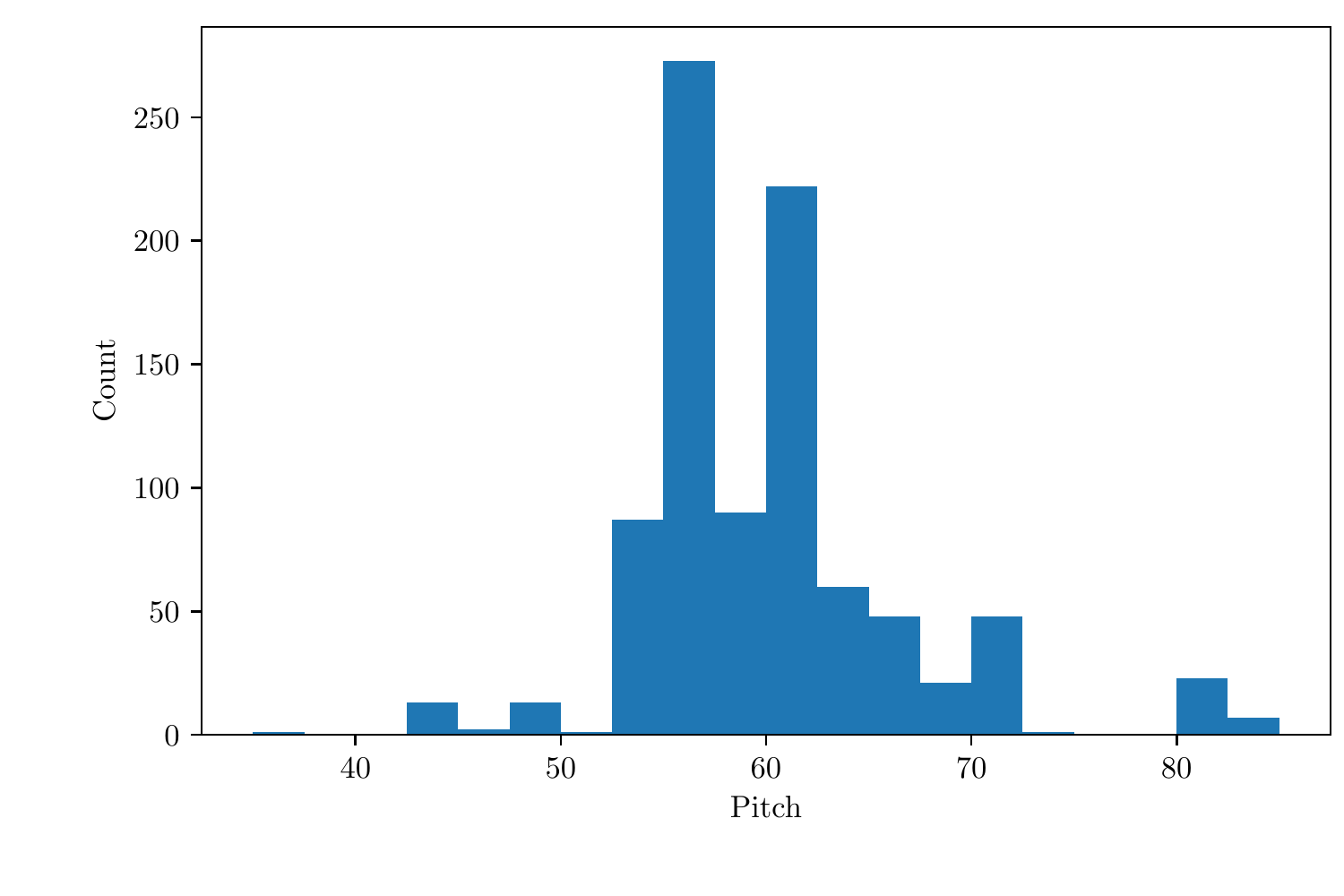}}}%
    \subfloat[$\tau= 0.9$]{{\includegraphics[height= 3.5cm, width=4cm]{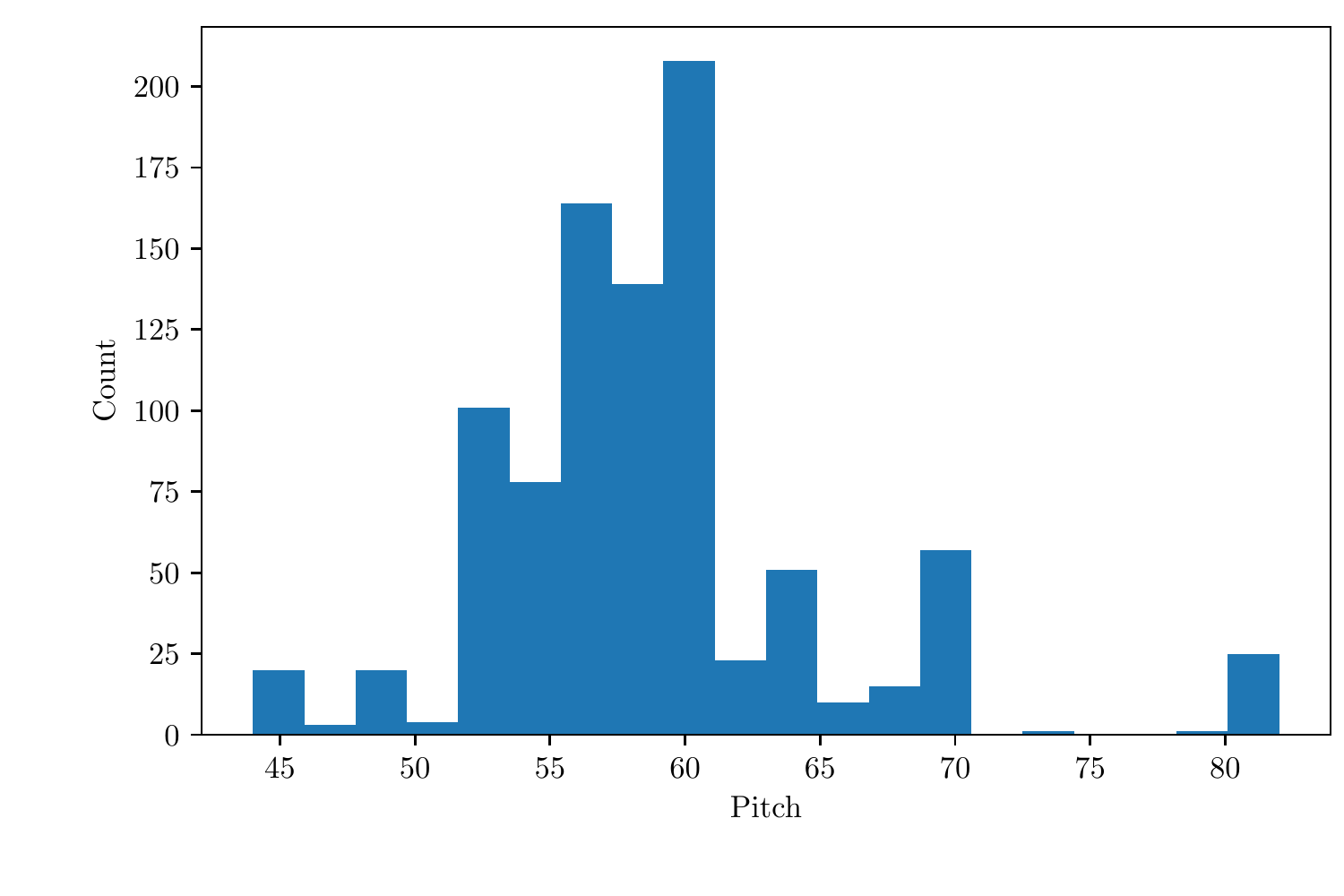}}}%
    \subfloat[$\tau = 1.0$]{{\includegraphics[height= 3.5cm, width=4cm]{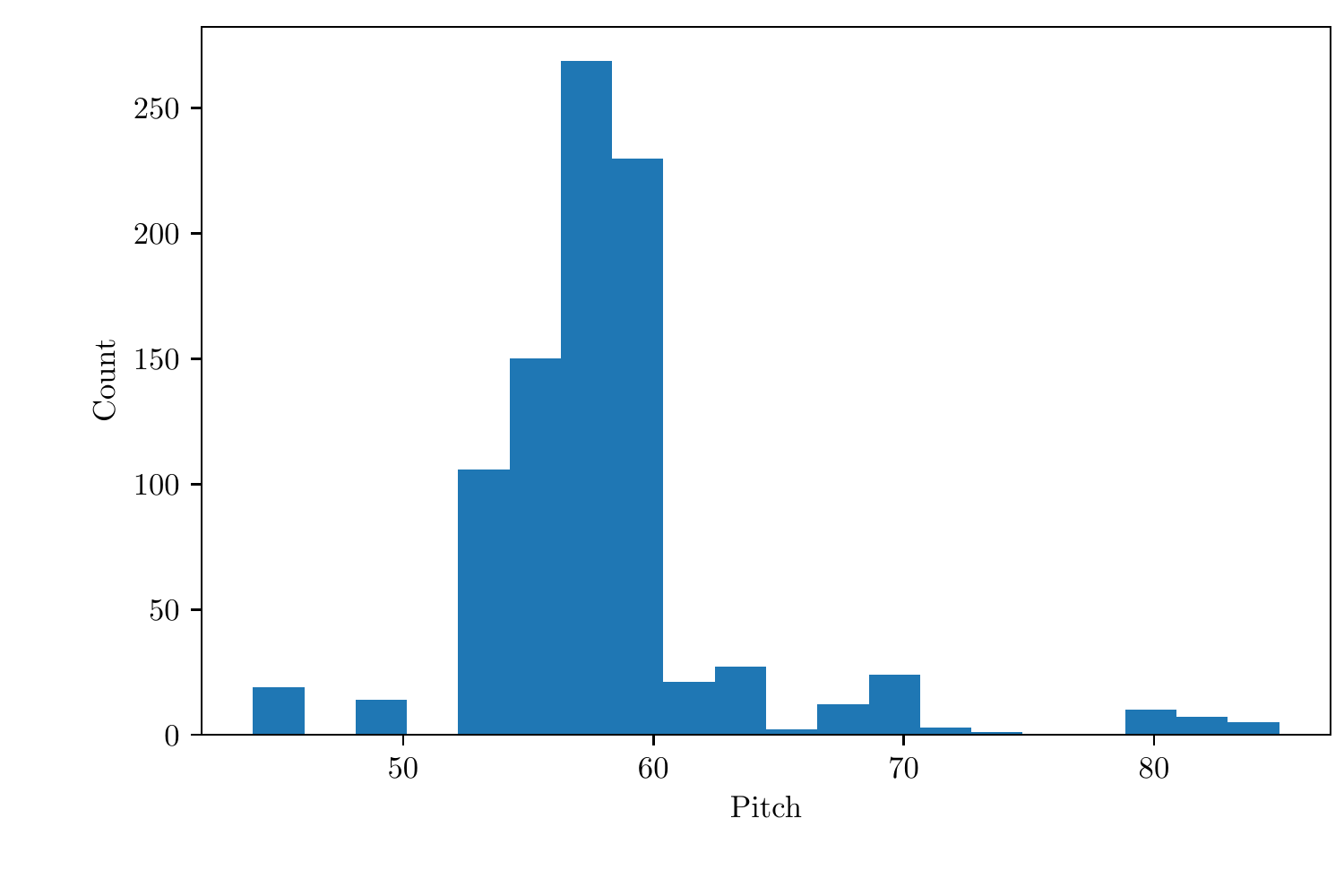}}}%
    \caption{Pitch histogram distribution of the generated lyrics predicted from melody decoder for greedy and temperature sampling methods with lyrics encoded by adding syllable and word embeddings (ASW).}%
    \label{fig:generated_pitch_syll_plus_word_emb}%
\end{figure*}


\begin{figure*}[htbp]
    \centering
    \subfloat[Greedy search]{{\includegraphics[height=3.5cm, width=4cm]{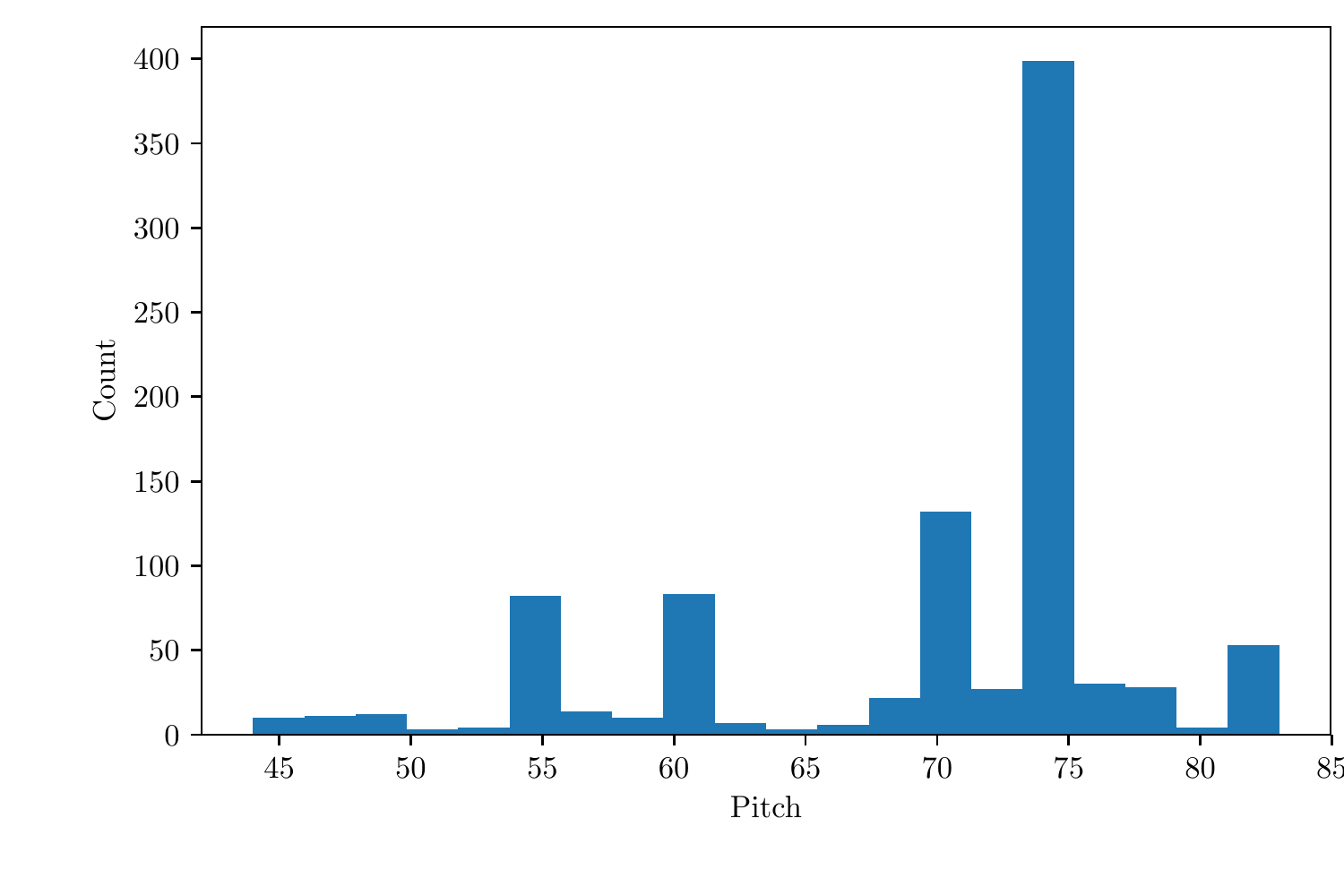}}}%
    \subfloat[$\tau = 0.5$]{{\includegraphics[height= 3.5cm, width=4cm]{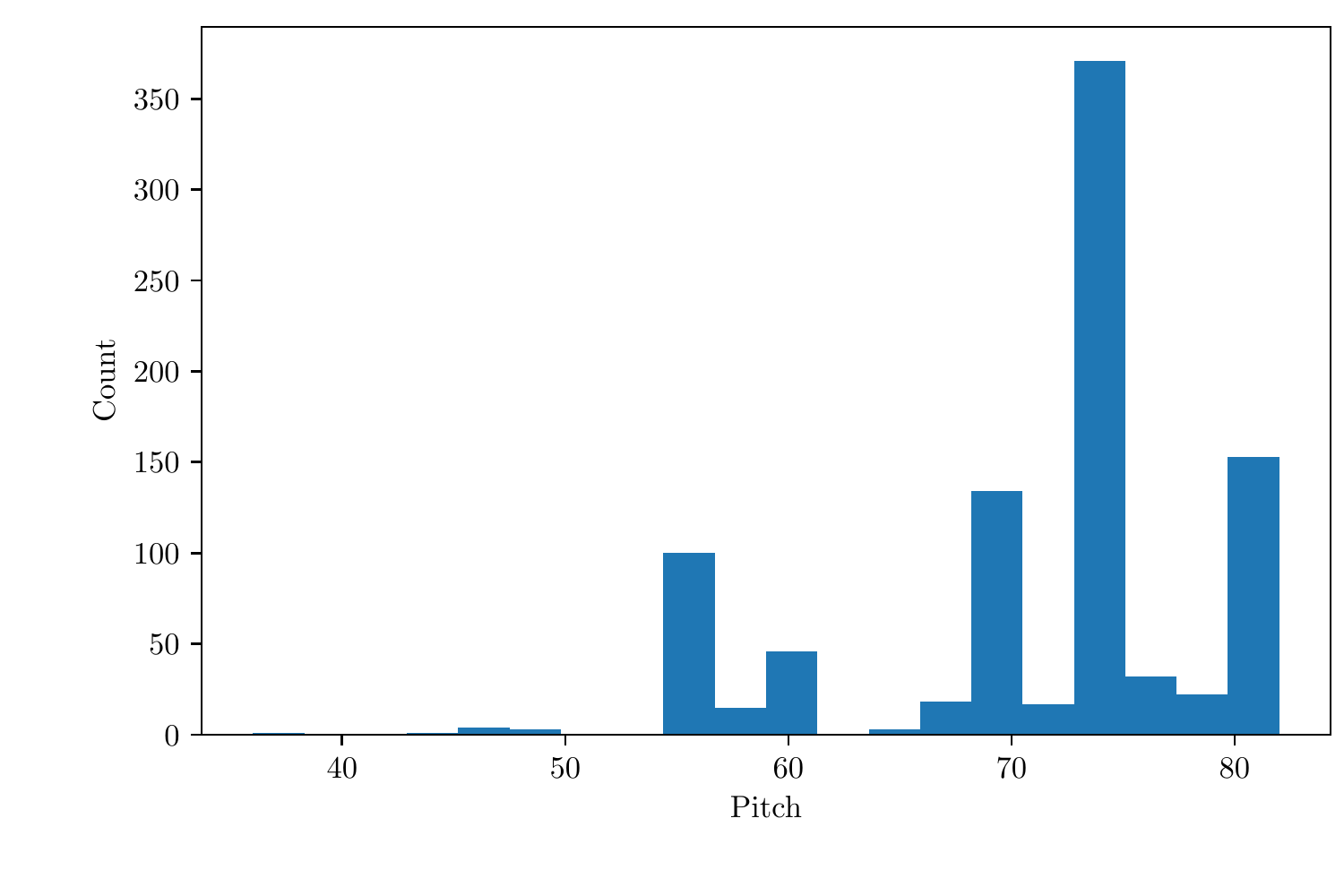}}}%
    \subfloat[$\tau = 0.6$]{{\includegraphics[height= 3.5cm, width=4cm]{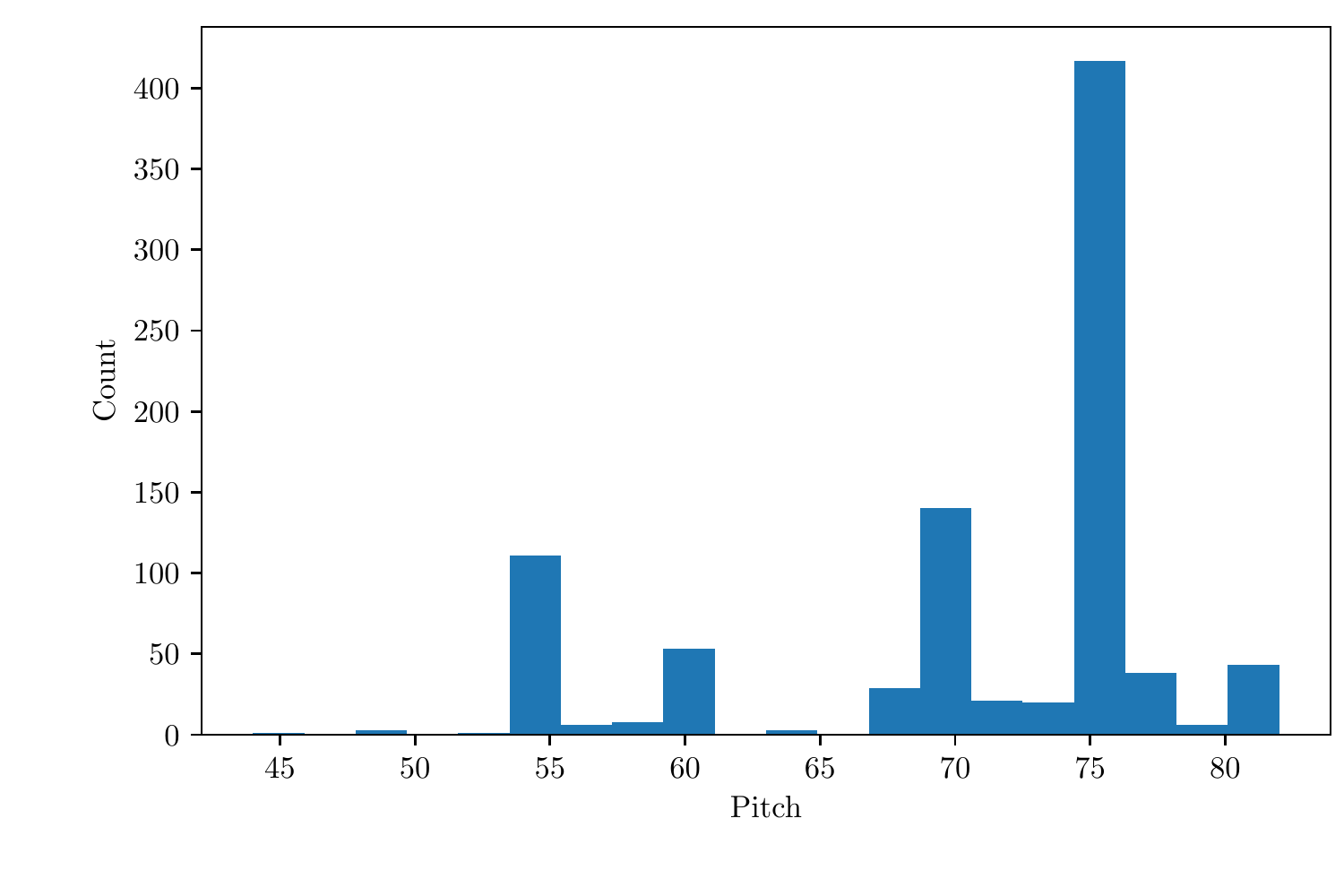}}}%
	\subfloat[$\tau = 0.7$]{{\includegraphics[height= 3.5cm, width=4cm]{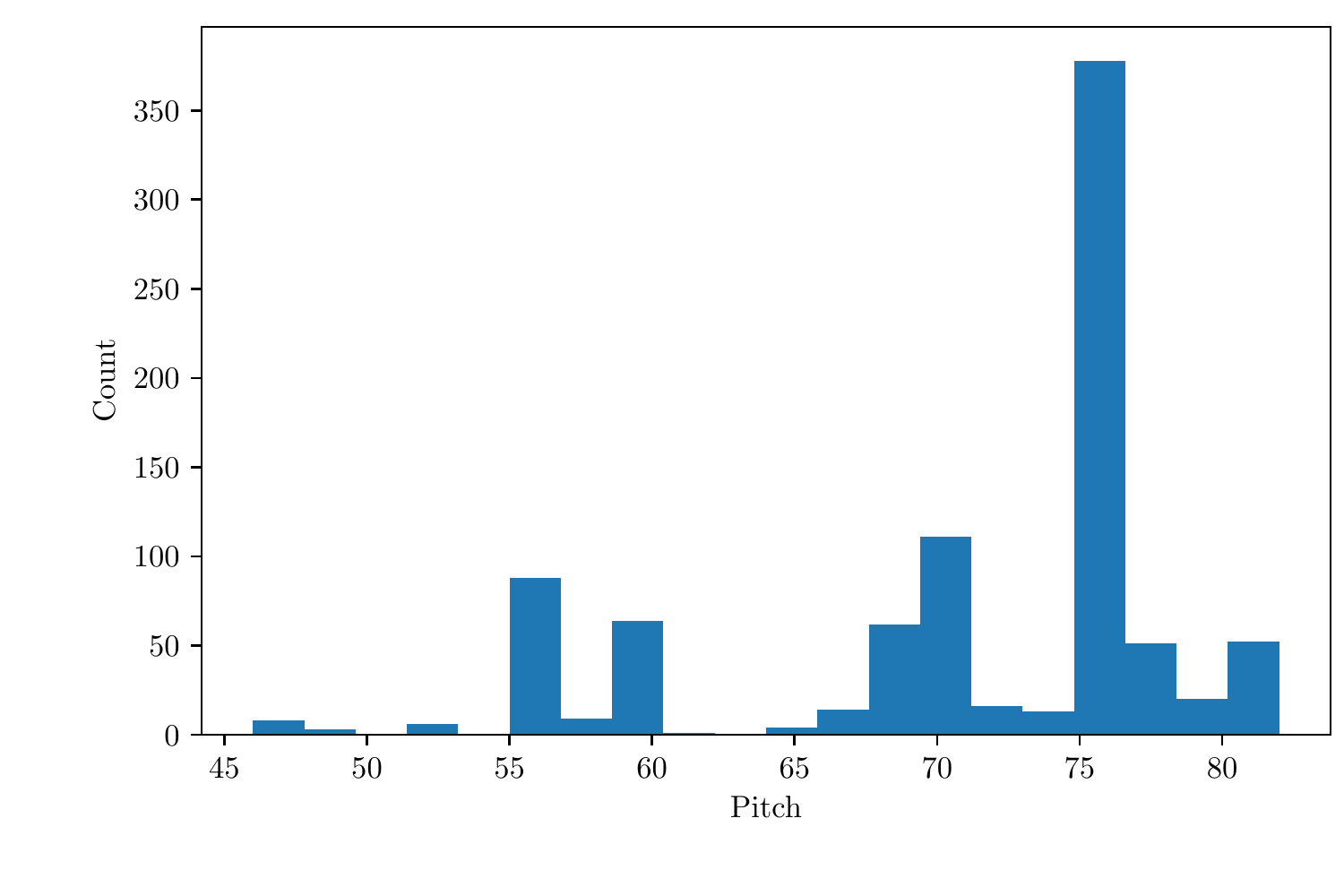}}}\\%
    \subfloat[$\tau = 0.8$]{{\includegraphics[height= 3.5cm, width=4cm]{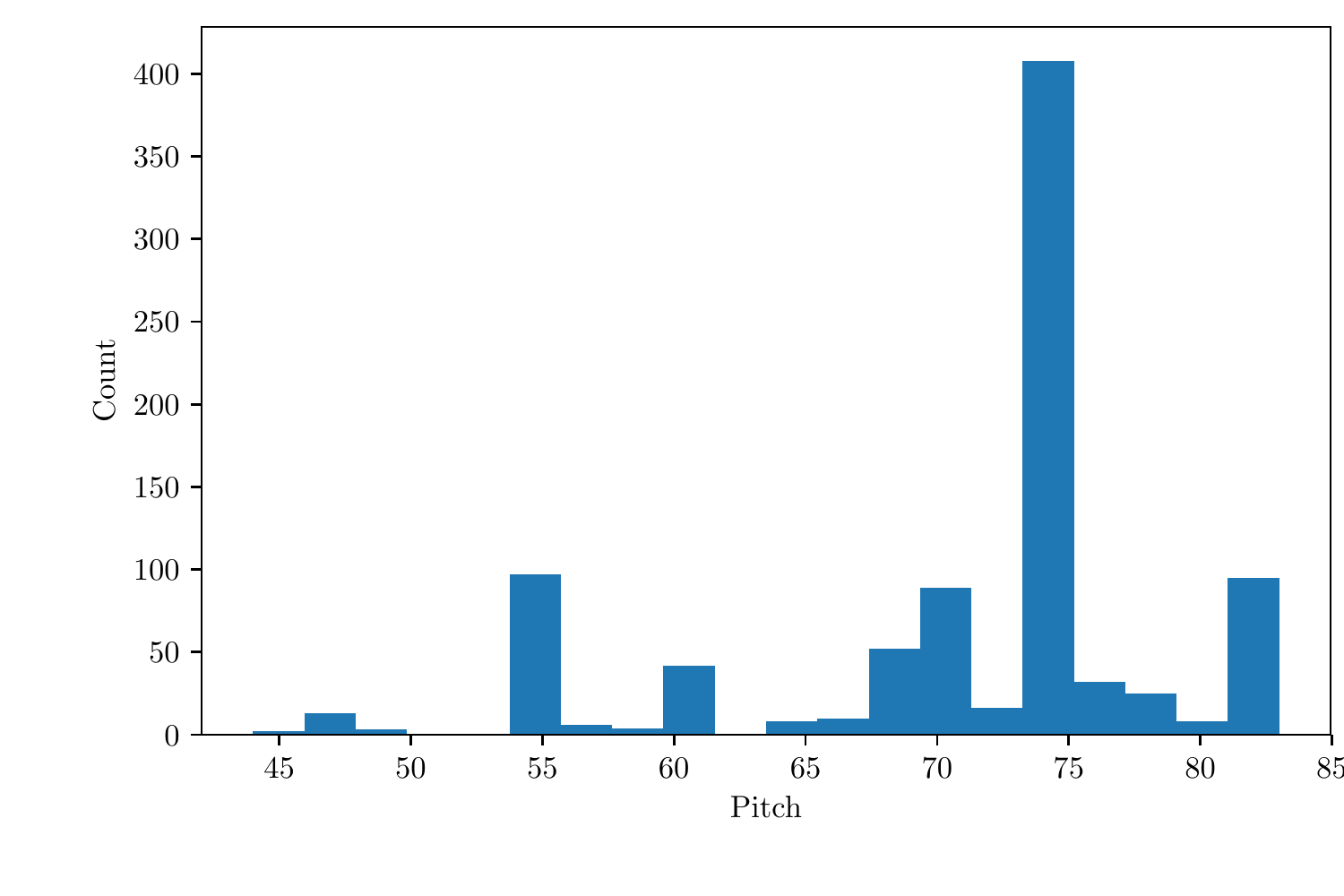}}}%
    \subfloat[$\tau= 0.9$]{{\includegraphics[height= 3.5cm, width=4cm]{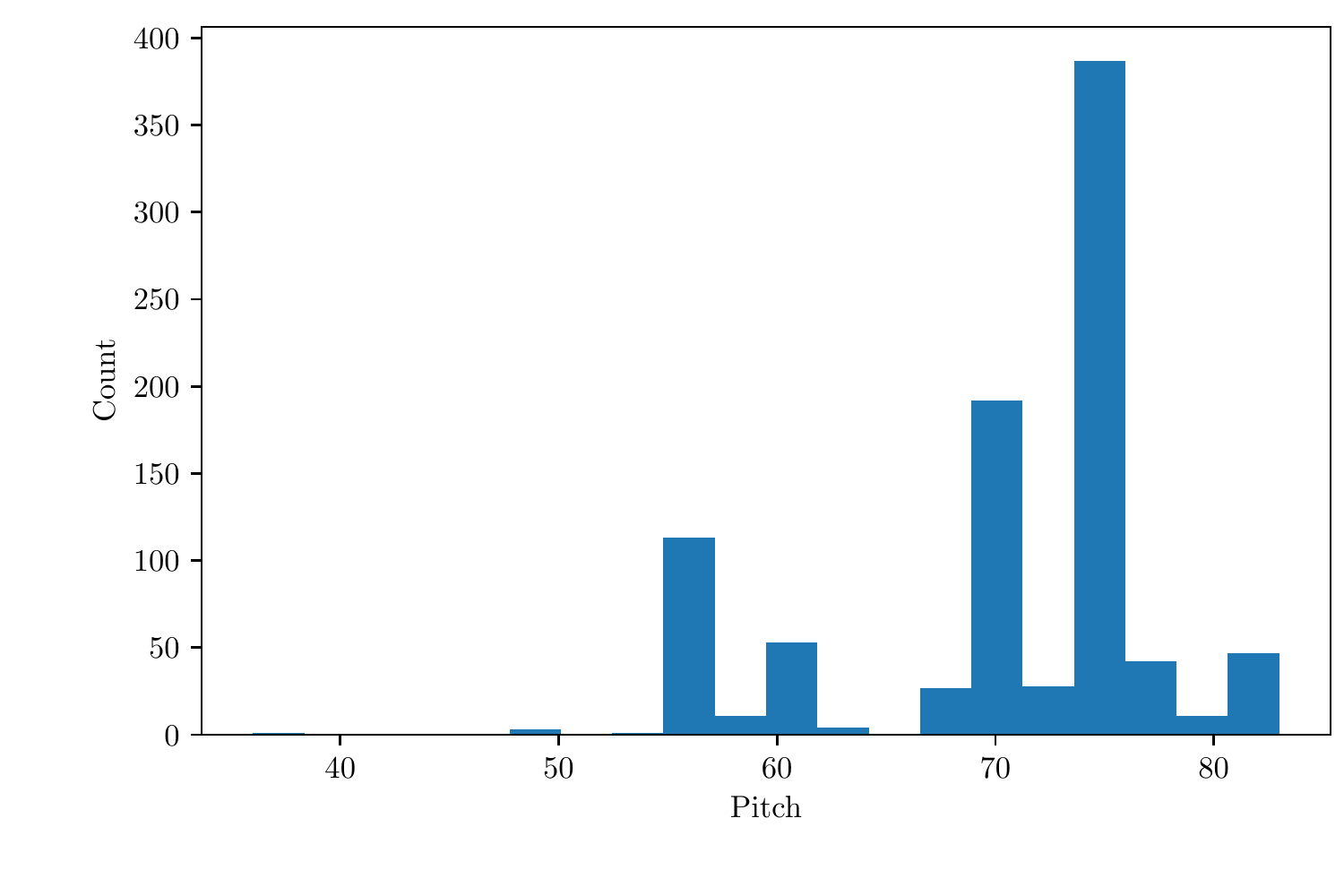}}}%
    \subfloat[$\tau = 1.0$]{{\includegraphics[height= 3.5cm, width=4cm]{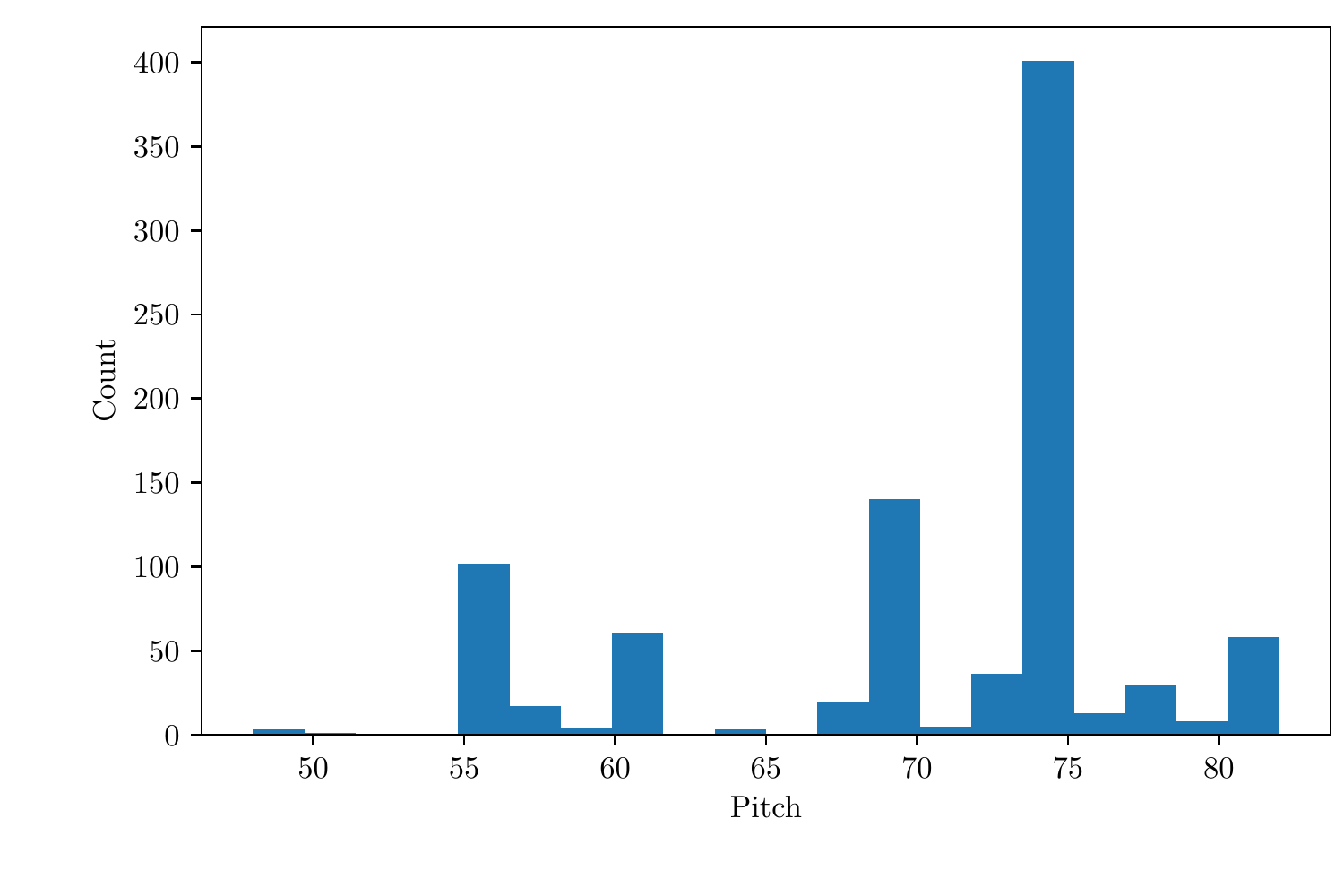}}}%
    \caption{Pitch histogram distribution of the generated lyrics predicted from melody decoder for greedy and temperature sampling methods with lyrics encoded by concatenating syllable, word and syllable projected word vector embedding (CSWP).}%
    \label{fig:generated_pitch_syll_word_proj_emb}%
\end{figure*}


\begin{figure*}[htbp]
    \centering
    \subfloat[Syllable embeddings (SE)]{{\includegraphics[height=3.5cm, width=4cm]{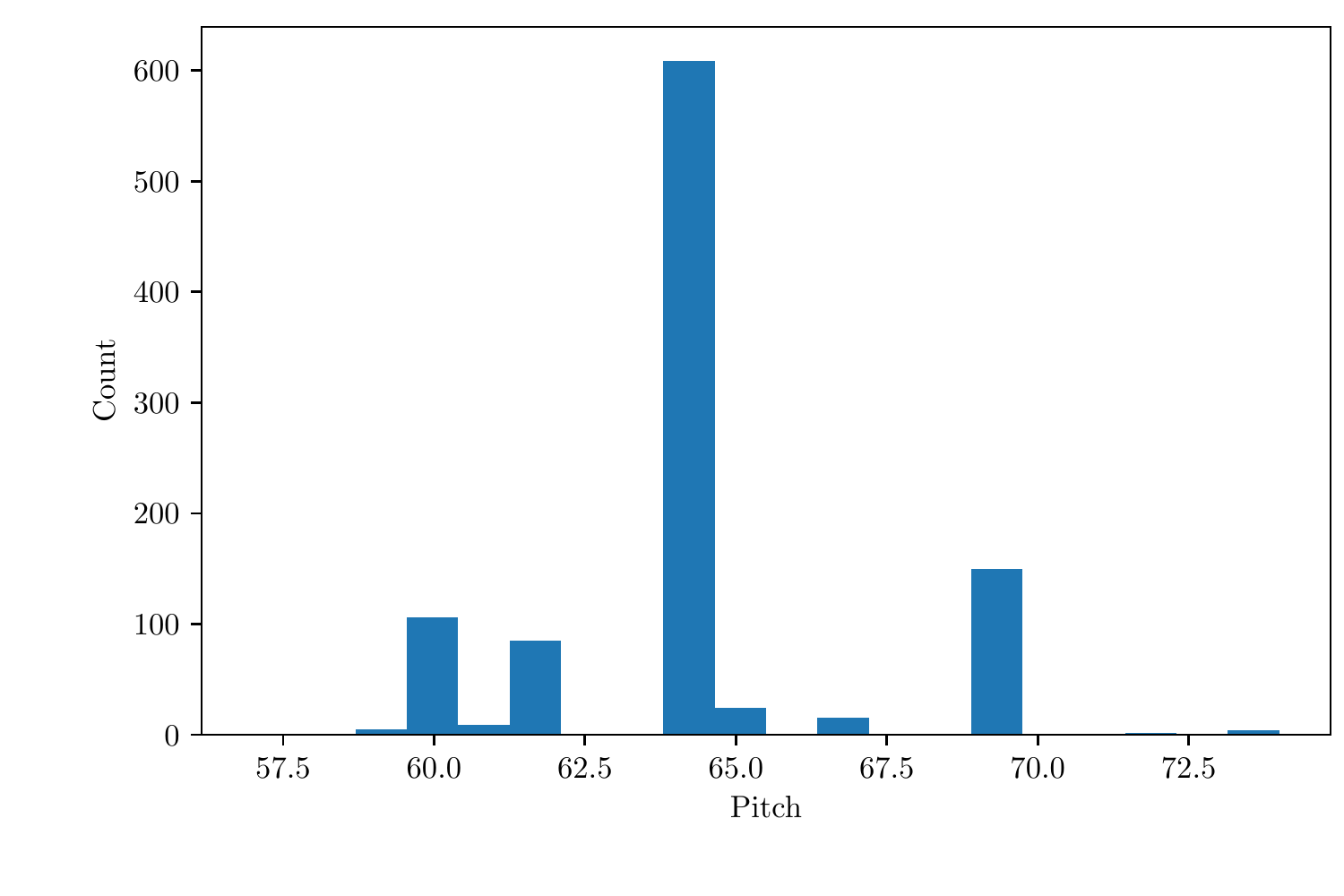}}}%
    \subfloat[syllable and word embedding concatenation (SWC)]{{\includegraphics[height= 3.5cm, width=4cm]{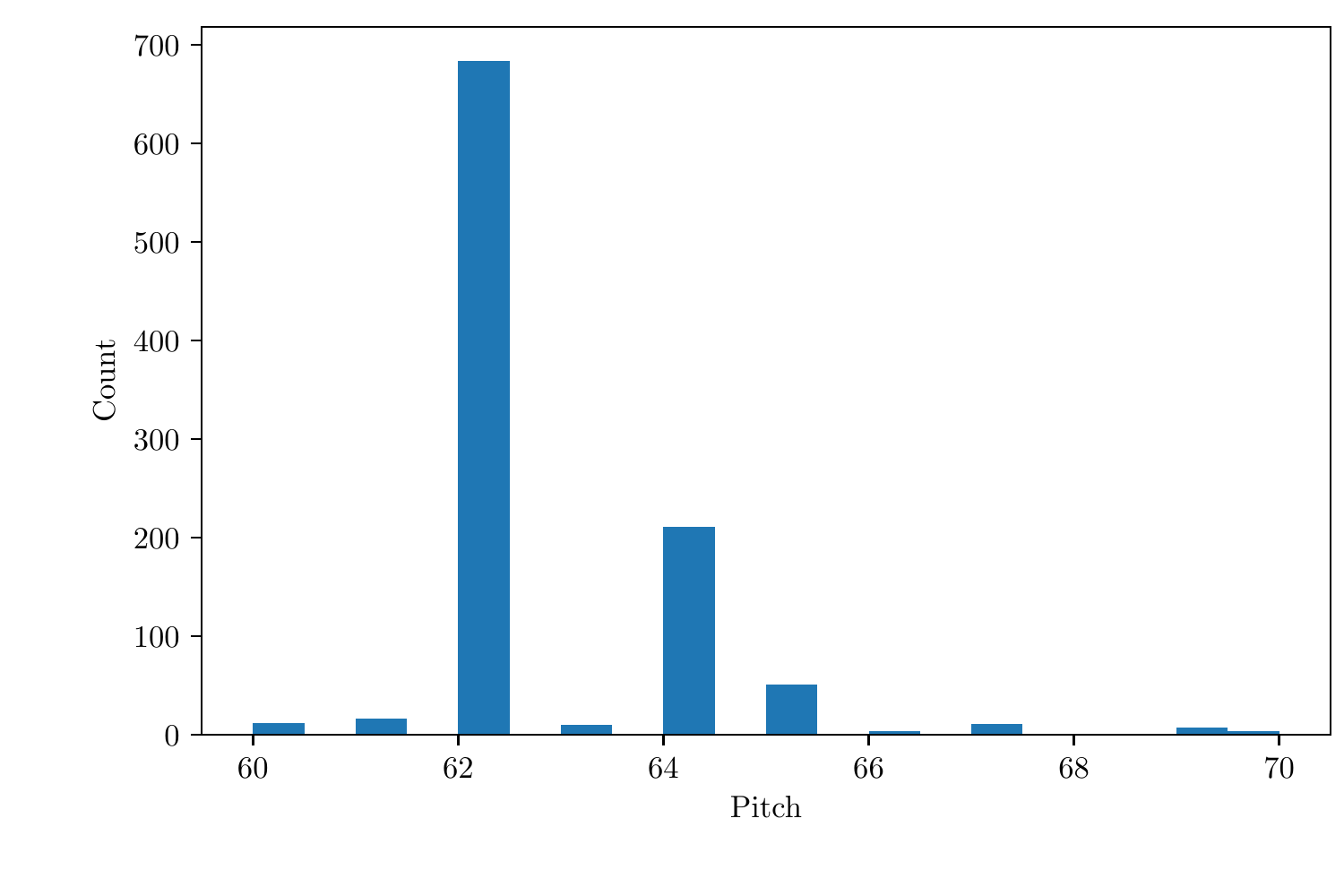}}}%
    \subfloat[addition of syllable and word embeddings (ASW)]{{\includegraphics[height= 3.5cm, width=4cm]{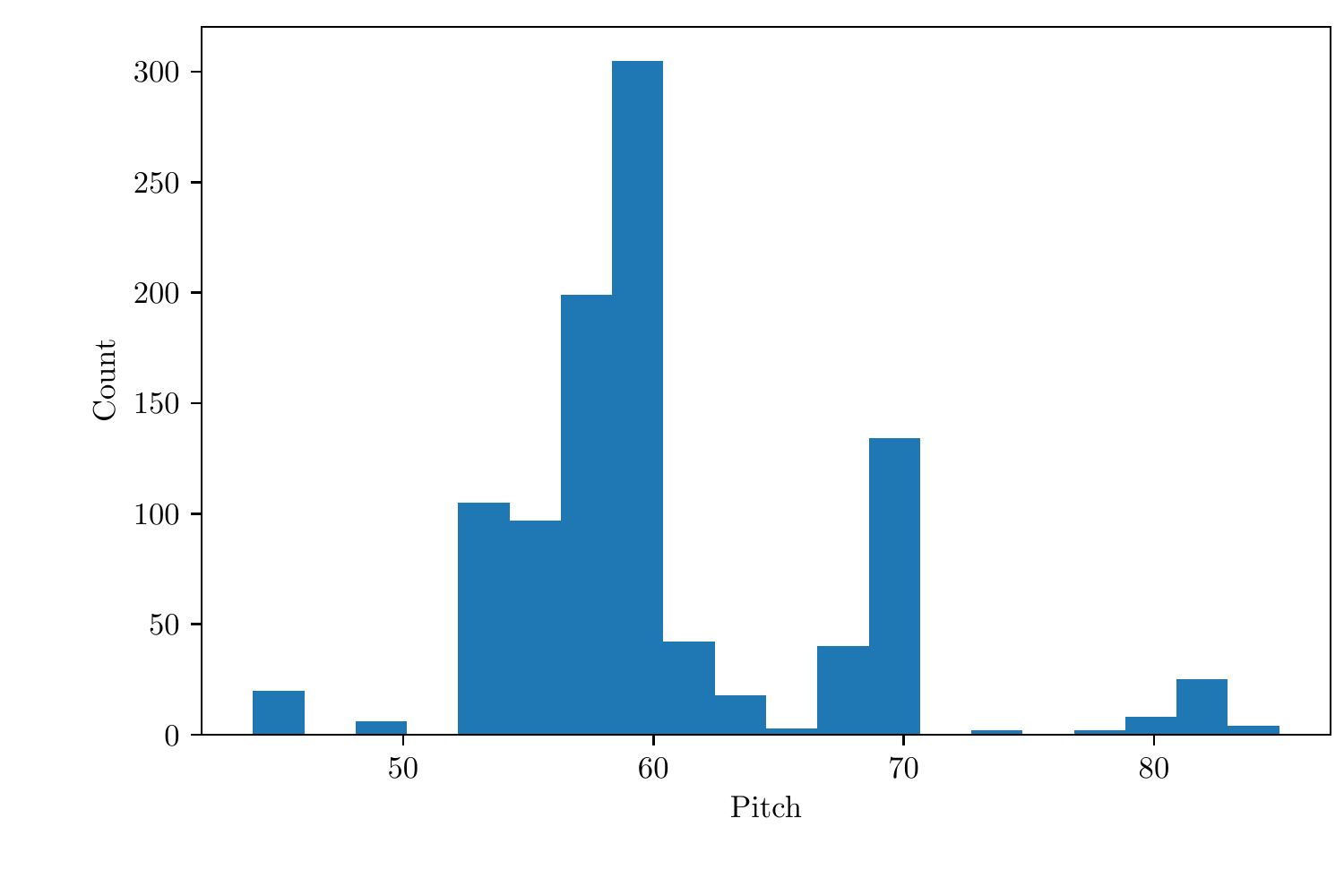}}}%
	\subfloat[concatenated syllable, word and syllable projected word vector
(CSWP)]{{\includegraphics[height= 3.5cm, width=4cm]{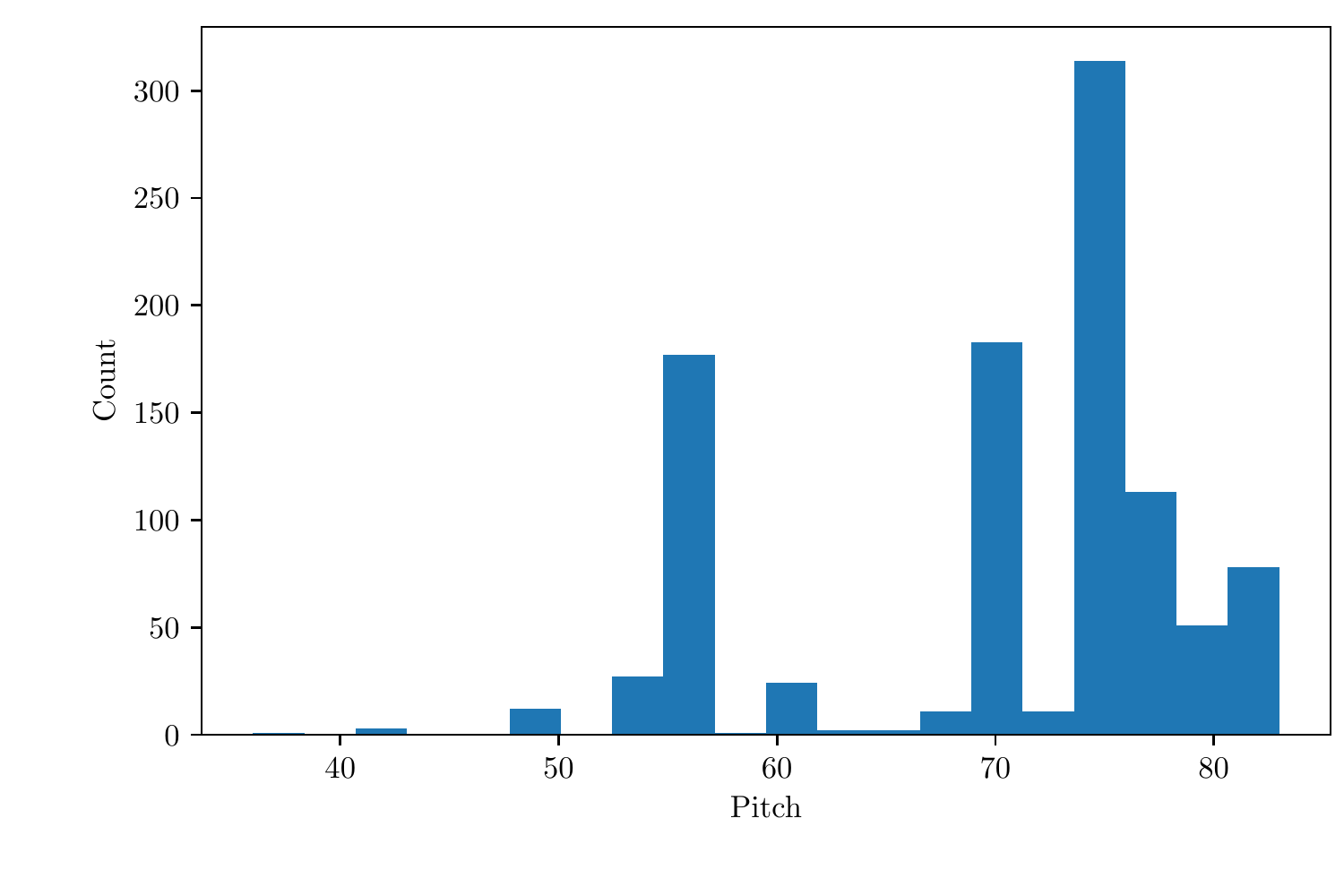}}}\\%
 \caption{Pitch histogram distribution of melody generated from the lyrics from test dataset for various lyrics embedding representations.}%
    \label{fig:original_lyrics_midi}%
\end{figure*}


\begin{figure*}[htbp]
        \centering
		\resizebox{16cm}{12cm}{\includegraphics{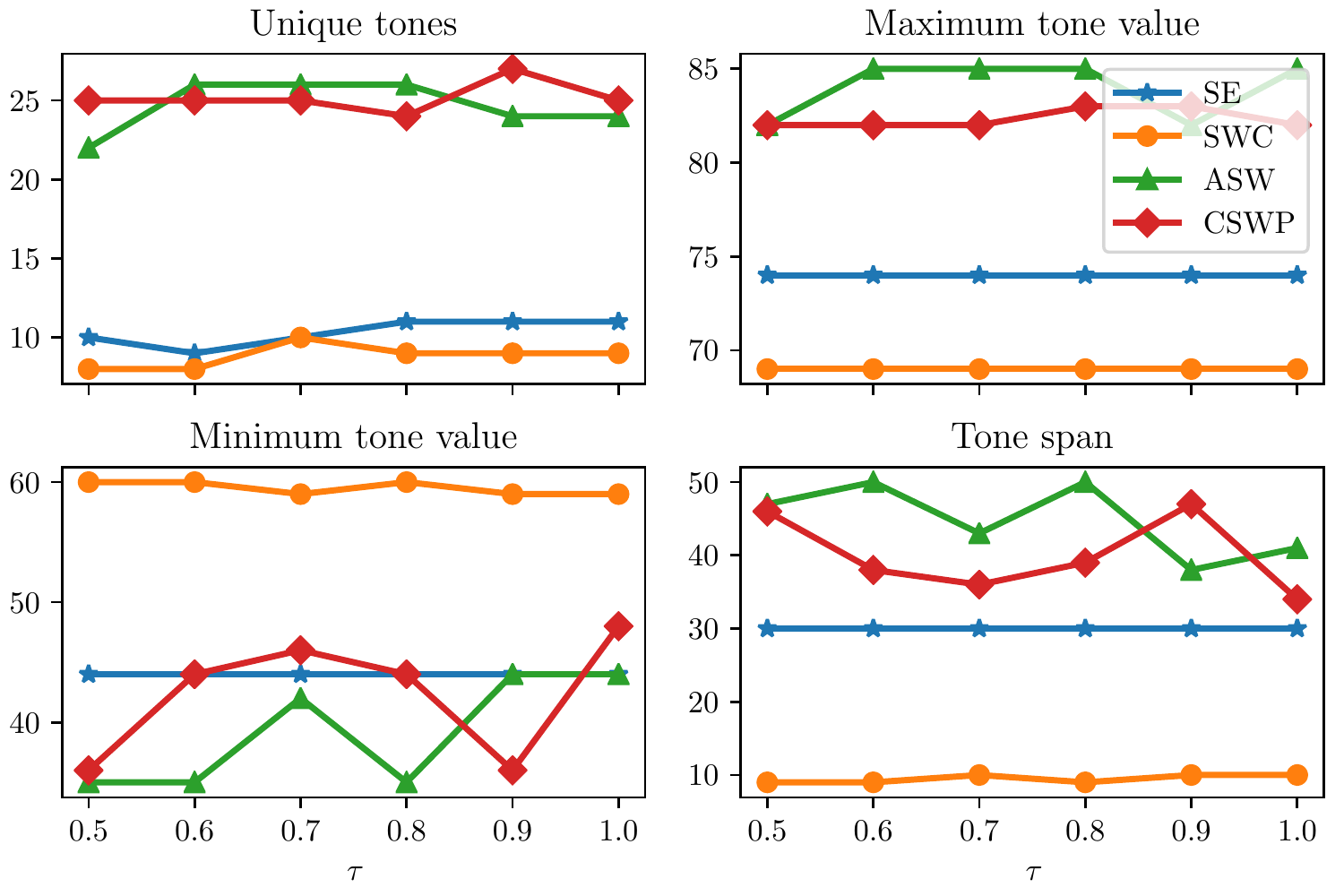}}  
        \caption{Objective measures for various $\tau$ values and embedding representations for the melody composed from generated lyrics.}
        \label{fig:comp_gen_lyrics_object}
 \end{figure*} 

\begin{figure}[htbp]
        \centering
		\resizebox{8cm}{4cm}{\includegraphics{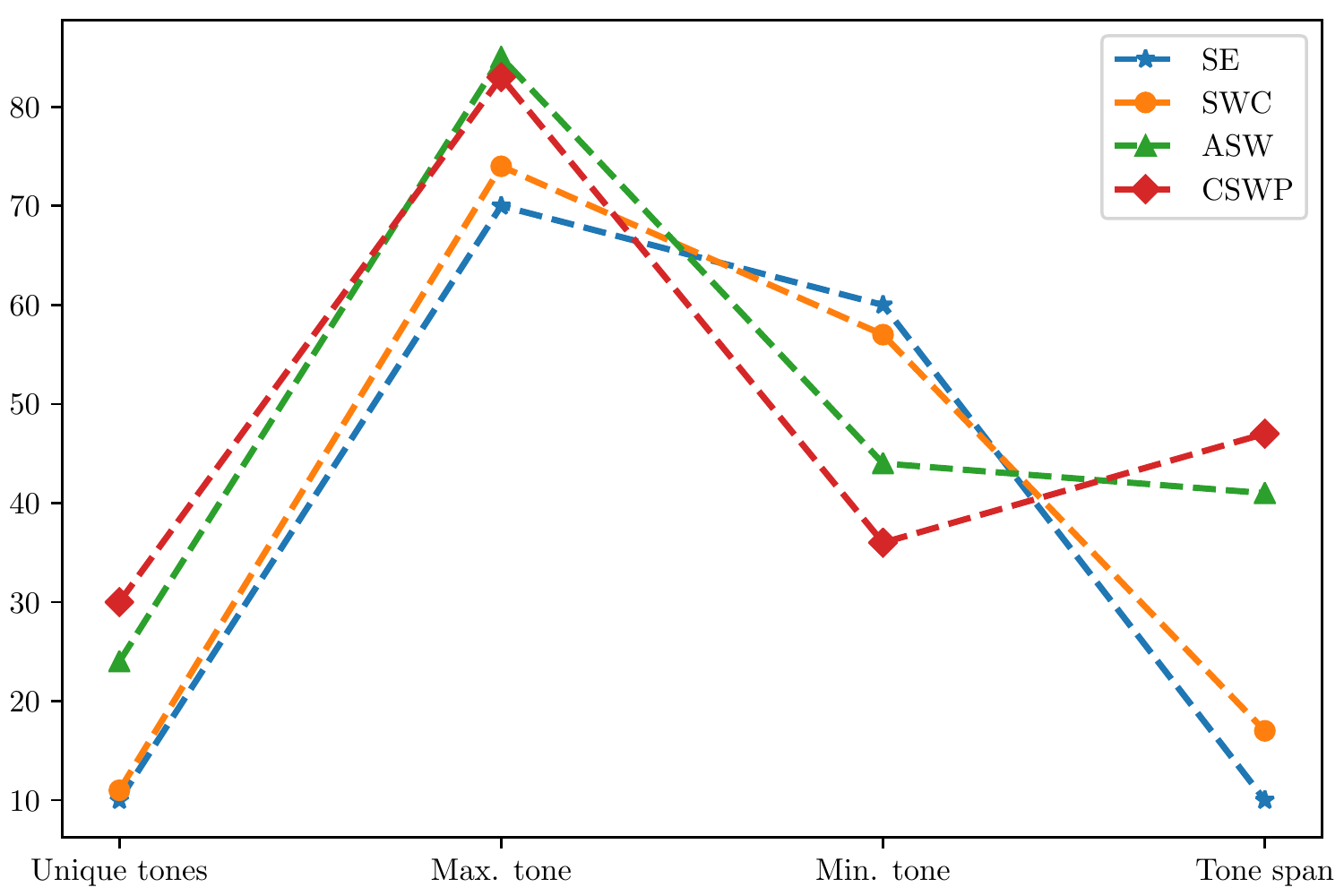}}  
        \caption{Objective measures for lyric embedding representations for test set lyrics.}
        \label{fig:comp_orig_lyrics_object}
 \end{figure} 

\begin{figure*}[htbp]
        \centering
		\resizebox{14cm}{8cm}{\includegraphics{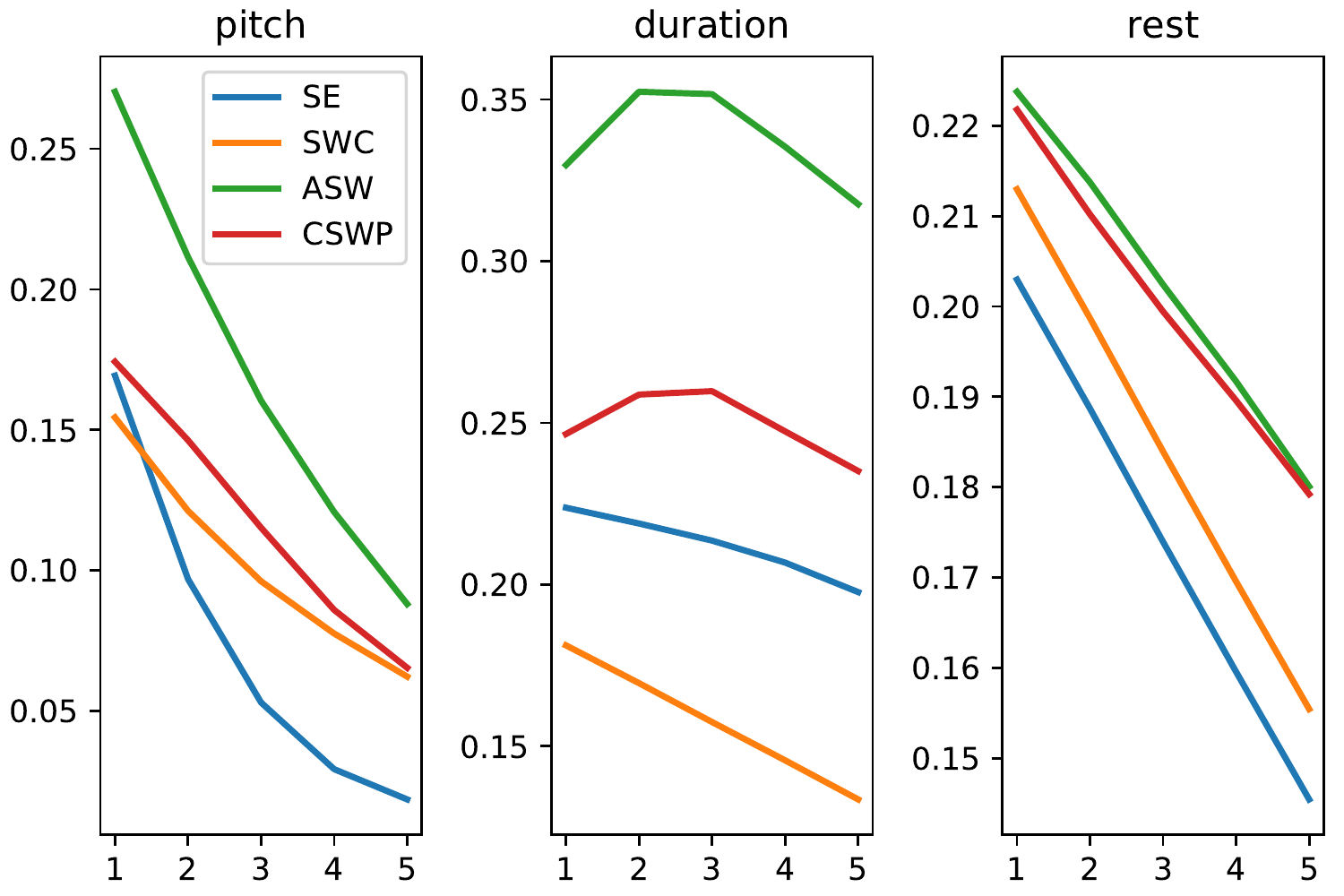}}  
        \caption{BLEU score evaluation of pitch, duration and rest melody attributes for SE, SWC, ASW and CSWP embedding representations. Abscissa represents the 1, 2, 3, 4, and 5 gram BLEU scores respectively.}
        \label{fig:bleu_score}
 \end{figure*} 

\begin{figure*}[htbp]
    \centering
    \subfloat[]{{\includegraphics[height=2cm, width=17cm]{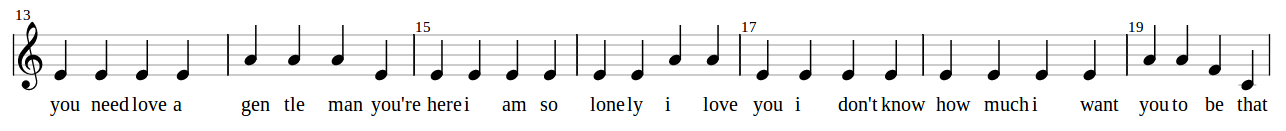}}}%
    \qquad
    \subfloat[]{{\includegraphics[height= 2cm, width=17cm]{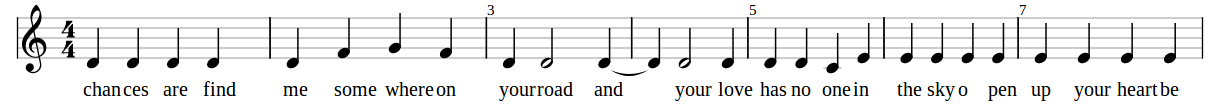}}}%
    \qquad
    \subfloat[]{{\includegraphics[height= 2cm, width=17cm]{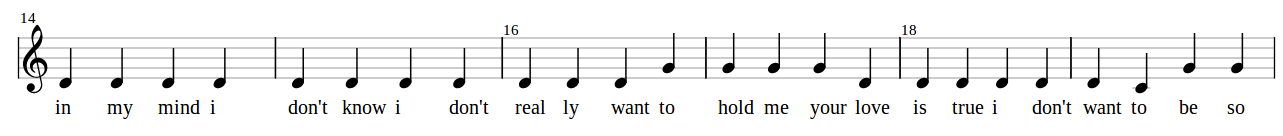}}}%
    \qquad
	\subfloat[]{{\includegraphics[height= 2cm, width=17cm]{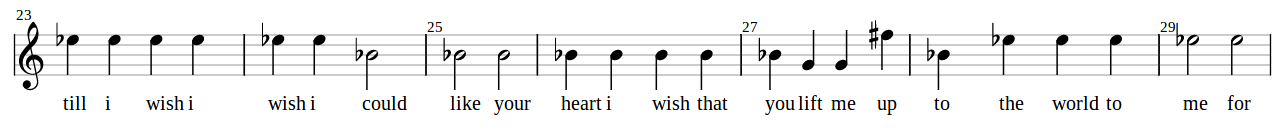}}}\\%

    \caption{Music sheet scores for generated lyrics-melody pairs for $\tau=0.8$ for SE, SWC, ASW and CSWP are show in Figs. a), b), c) and d) respectively.}%
    \label{fig:generated_lyrics_melody_pairs}%
\end{figure*}


\section{Dataset Description} 
\label{sec:dataset}

The dataset used in our experiments initially created in~\cite{yu2019conditional}, which come from two different sources: a) the "LMD-full dataset" of the Lakh MIDI Dataset v0.1 \footnote{\url{https://colinraffel.com/projects/lmd/}} and a dataset found on a reddit thread called "the largest midi collection on the internet"\footnote{\url{https://https://www.reddit.com/r/datasets/comments/3akhxy/the_largest_midi_collection_on_the_internet/}}. The LMD-full dataset contains a total of 176,581 different MIDI files but, after taking only MIDI files with sufficient English lyrics,
only 7,998 files could be used. The largest midi collection on the internet dataset contains 130,000 different MIDI files but only 4,025 with enough English lyrics. These two datasets were merged together to get a total of 12,023 MIDI files. For our experiments, we parsed the dataset into three different formats: a) The syllable level: This format is the lowest level that pair together every notes and the corresponding syllable and its attributes. b) The word level: This format regroups every notes of a word and gives the attributes of every syllables that makes the word. c) The Sentence level: Similarly, this format put together every notes that forms a syllable (or in most case, a lyric line) and its corresponding attributes. Each syllable is represented with discrete attributes: a) The Pitch of the note: In music, the pitch is what decide of how the note should be played. We use Midi note number as an unit of pitch, it can take any integer value between 0 and 127. b) The duration of the note: The duration of the note in number of staves. It can be a quarter note, a half note, a whole note or more. The exhaustive set of values it can take in our parsing is [0.125, 0.25, 0.5, 0.75, 1, 1.5, 2, 4, 8, 16, 32]. and c) The duration of the rest before the note: This value can take the same numerical values as the duration but it can also be null (or zero).

The algorithm to parse the MIDI files to extract the lyric syllables and the corresponding melody attributes is given in Algorithm 1. For the word and sentence level, the algorithm is extremely similar, with just the part about recognizing what is a word and what is a sentence that is added.

\begin{algorithm}[htbp]
\SetAlgoLined
 open a midi file\;
 $i\leftarrow 0$\;
 $TrackNumber\leftarrow 0$\;
 \While{note's timestamp and first lyric's timestamp not almost equal}{
   \tcp{The first 5 notes of every tracks are checked until there is an equality}
  \eIf{$i=5$}{
  
  $i\leftarrow 0$\;
  $TrackNumber\leftarrow TrackNumber + 1$\;
  }{
  $i\leftarrow i + 1$\;}
  }
  (\tcp*[f]{Here the TrackNumber is found})
  Go on i-th first note of the found Track\;
  $ListNote\leftarrow [\ ]$\;
  \tcp{The iteration and parsing of the song begins here}
  \While{The song and the lyrics has not reached their end}{
  \eIf{The next note and lyric's timestamp match}{
  $ListNote\leftarrow ListNote + CurrentNote$\;
  Iterate to the next note and next lyric\;
  }{
  \eIf{The note is a lot before the lyrics's start}{
  Iterate to the next note\;}{
  Iterate to the next lyric\;}}
  }
  close the file\;
 \caption{Syllable level parsing}
\end{algorithm}


The pitch distribution of the dataset is shown in Fig.~\ref{fig:dataset_pitch_dist}. From distribution plot in Fig.~\ref{fig:dataset_pitch_dist}, we can observe that most of our songs have pitch range between 60 and 80 midi values. With mean pitch value of approximately 75. The histogram distribution of the duration of all our notes is shown in Fig.~\ref{fig:dataset_duration}.  From Fig~\ref{fig:dataset_duration}, we can observe that most of our notes are quarter notes followed by half notes and other notes. Similarly, we can observe the distribution of the rest attribute of our dataset in Fig.~\ref{fig:dataset_rest}. From the rest distribution in Fig.~\ref{fig:dataset_rest}, we can observe that almost 90\% of our songs have zero rest between the notes. The only significant rest attribute between the notes is quarter, half and full duration rests. We also plot the histogram distribution of the pitch interval in the dataset which is show in Fig.~\ref{fig:dataset_pitch_interval}. From Fig.~\ref{fig:dataset_pitch_interval}, we can observe that the pitch interval distribution can be approximated with a Gaussian with zero mean and standard deviation of approximately 5. i.e., most of our consecutive note sequences have smooth pitch changes and they span form 0 to 5 midi pitch values. This also signifies that we have a sufficient data which can be modeled to fit the distribution. The pitch range distribution of all songs in the dataset is shown Fig.~\ref{fig:dataset_pitch_range}. We can observe that an average, we have approximately 15 tones gap between the minimum and the maximum note value in each song with the variation of $\pm 5$ tones. 

\begin{table*}[htbp]
\centering
\caption{Seed lyrics and the generate full length songs ($\tau$ = 0.6).}
\label{table:seed_lyrics}
\begin{tabular}{|c|c|}
\hline
\textbf{Seed lyrics}                                     & \textbf{Generated lyrics}                                                                                                                                                                                                                                                                                                                                                                                                                                                                                     \\ \hline
i never needed you to be strong                          & \begin{tabular}[c]{@{}c@{}}i never needed you to be strong i never\\  saw you you any time i could we be got\\  i believe you i will always will be the one \\ i love you if you know it's just a girl i can stop\\  the music in the night you can make me \\ feel the way now i know you will see something \\ that i need i feel alone i wanna be alone with \\ out a love to be a part how am i supposed to\end{tabular}                                 \\ \hline
chances are find me some where on your road              & \begin{tabular}[c]{@{}c@{}}chances are find me some where on your road \\ so the morning comes before i believe katy she\\  walks the one is me pa pa pa to three years \\ to rock me pretending let me be the one we fly \\ i will be there when we have to let the music \\ we belong together i don't know now i know\\  i know but i know i know i need i got\\  to believe i would n't cry i would n't let me\end{tabular}  \\ \hline
our world is miles of endless roads                      & \begin{tabular}[c]{@{}c@{}}our world is miles of end less roads a drummer \\ walks we always be mellow nun y happy \\ it's i never mind in love and in your heart\\  that's that ing i don't need to know if are i find \\ your self i can never be my baby but i\\  can't believe that i'm not the one i'm waiting for \\ the rain on the rain once makes me cry on t\\ he rain on the rain looking out that i still have\end{tabular}                          \\ \hline
\end{tabular}
\end{table*}

\begin{figure*}[htbp]
        \centering
		\resizebox{18cm}{4cm}{\includegraphics{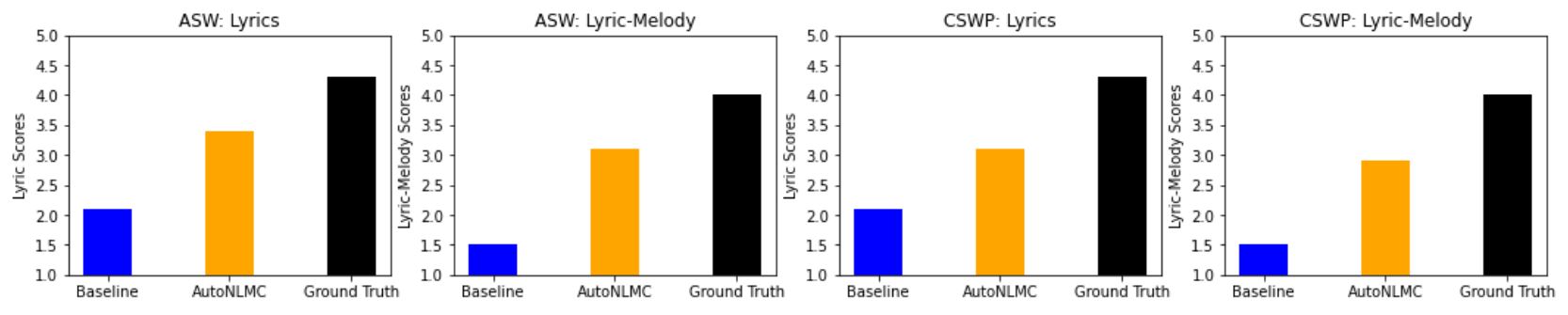}}  
        \caption{Subjective evaluation results for lyrics and lyric-melody pairs. Lyrics and melodies are generated from ASW and CSWP of the proposed AutoNLMC for $\tau= 0.6$.}
        \label{fig:subjective_score}
 \end{figure*} 

\section{Experiments}
\label{sec:experiments}

We have experimented with greedy search and temperature sampling as discussed in subsection~\ref{subsec:inference} methods to generate the lyrics from the lyrics prediction model. We generate 10 full length songs by constraining the number of syllables generated in each song to be 100 syllable length. We use ten variable length seed lyrics i.e., lyrics to start with generating full song lyrics form the test dataset. Few examples of seed lyrics used to generate the full length song lyrics and the generated lyrics are shown in Table~\ref{table:seed_lyrics}. 
We evaluate the quality of the generated lyrics qualitatively by comparing the distributions of the melody composed from the generated lyrics and the dataset melody distributions. That is, for all the generated lyrics, we compose the melody by the melody composition model and compare the individual melody attribute distributions: pitch, duration and rest with the dataset distributions described in section~\ref{sec:dataset}. Also, as discussed in section~\ref{sec:feat_repre}, we represent the lyrics tokens in various encoding forms and train the proposed lyrics generator and melody decoder model one for each representation. The histogram distribution of the pitch decoded from the generated lyrics for greedy search and various temperature values $\tau$ of temperature sampling is shown in Figs.~\ref{fig:generated_pitch_syll_emb},~\ref{fig:generated_pitch_syll_word_concat_emb},~\ref{fig:generated_pitch_syll_plus_word_emb},~\ref{fig:generated_pitch_syll_word_proj_emb} for lyrics encoded with syllable embedding (SE), addition of syllable and word vector (ASW), syllable and word
embedding concatenation (SWC) and concatenated syllable, word and syllable projected word vector
(CSWP) respectively. From Fig.~\ref{fig:generated_pitch_syll_emb}, we can observe that the model trained with syllable embeddings alone shows quite interesting distribution similar to the dataset. But a careful observation reveals that the predicted pitch is biased towards the lower pitch range. This signifies that the syllables alone cannot capture the whole dataset distribution. The model trained by concatenating syllable and word embeddings with an anticipation to improve the distribution range does not succeeded in capturing underlying distribution shown in Fig.~\ref{fig:generated_pitch_syll_word_concat_emb}. Instead, it confines the distribution to a small range thus the model helps in preventing accidental abrupt change in pitch between notes. Though this type of model is quite stable in generating stable lyrics with no sudden jumps in pitch values between notes, it will be less interesting because the generated songs will have very low pitch span and less number of unique notes results in mostly monotonous melody for any given lyrics. We found that the model trained with the sum of syllable and word embedding vectors (ASW) shows interesting results with dense distribution round the mean pitch value of the dataset distribution. Unlike the distributions of models trained with SE and CSWP, ASW shows a more bell like distribution similar to dataset distribution results in picking other pitches with more likelihood. The pitch distribution for the model trained with CSWP embedding vectors is shown in Fig.~\ref{fig:generated_pitch_syll_word_proj_emb}. We have concatenated the projection syllable vector on world vector in order to give the wordness to the syllable vector since syllable and word embedding models trained separately. The pitch distribution of the model trained with CSWP captures the distribution range similar to the dataset distribution but the model most likely predicts the mean pitch of the dataset distribution resulting in predicting average melody of the dataset. We can conclude that ASW and CSWP embeddings are most likely suitable embeddings for generating meaningful melodies qualitatively. We can also note that greedy search can generate at most one lyric for a seed lyric but temperature based sampling can generate original lyric and melody pair for each $\tau$. It should be noted that for various $\tau$, the range of the distribution remains same with slight changes in the shape of the distribution that signifies that the lyrics generated by temperature sampling is not random and we can generate novel lyrics-melody pairs with each value of $\tau$. We also noted that the $\tau \in [0.5 - 1.0]$ is the best range to generate the lyrics-melody pairs which follows dataset distribution. Further, we can also conclude that ASW representation of lyrics can learn the correlation between the generated lyrics and melody in an effective manner. Since we can also generate the melody by feeding the original human generated lyrics to our model, we have composed the melody for 10 randomly chosen full length lyrics. 
The pitch histogram distribution shown in Fig.~\ref{fig:original_lyrics_midi} also confirms that the ASW and CSWP lyric embedding representations are better representations for capturing the lyric-melody correlations. We also investigate the musical note attribute related objective measures proposed in~\cite{mogren2016c}: 1) unique tones, which measures the total number of unique tones generated by the model, 2) maximum tone value, which gives the highest tone value generated, 3) minimum tone value, which represents the minimum tone value generated and 4) tone span, which is the range of the pitch values present in the generated melody to validate the claim. Fig.~\ref{fig:comp_gen_lyrics_object} shows the tone related objective measures for the melody composed from generated lyrics for various $\tau$ values and lyric embedding representations. From the left top Fig.~\ref{fig:comp_gen_lyrics_object}, we can observe that ASW and CSWP embedding representations generate a large number of unique tones compare to other representations. Similarly, ASW and CSWP representations are capable of generating very low and very high tone values which is consistent with the dataset distribution which can be observed from left bottom and right top of Fig.~\ref{fig:comp_gen_lyrics_object}. The tone span plotted in the bottom right of Fig.~\ref{fig:comp_gen_lyrics_object} also shows that ASW and CSWP representations are capable of generating larger tone span melodies compared to other representations. The tone related objective measures in Fig.~\ref{fig:comp_gen_lyrics_object} also reveal that irrespective of $\tau$ value, the objective measures remains constant for SE and SWC which indicated that these representations are not capable of generating expressive melodies for different values of $\tau$ even though generated lyrics are quite different for each $\tau$ value. Whereas ASW and CSWP are capable of generating interesting tones for various values of $\tau$. The objective measures for the pitch generated from the test set lyrics is shown in Fig.~\ref{fig:comp_gen_lyrics_object} also confirms that ASW and CSWP are indeed good representations for generating melodies. In~\cite{bao2018neural}, authors proposed to use BLEU (bilingual evaluation understudy)~\cite{papineni2002bleu} as a metric for evaluating the predicted pitch from lyrics. We also evaluate our model with BLEU score for predicted pitch, duration and rest. BLEU is a single real number value to evaluate the translation quality of the model from input to output. Here, we compute the BLEU scores for the melodies generated from the test set lyrics and corresponding ground truth melodies. Higher BLEU score indicates better correlation between the generated and ground truth melodies. The 1-gram, 2-gram, 3-gram, 4-gram and 5-gram BLEU scores of pitch, duration and rest melody attributes for each embedding representations are shown in Fig.~\ref{fig:bleu_score}. From Fig.~\ref{fig:bleu_score}, we can observe that ASW and CSWP representations out performs other representations. 
Few samples of music sheet scores for the generated lyrics and melody pairs by the proposed AutoNLMC is shown in Fig.~\ref{fig:generated_lyrics_melody_pairs}. More samples can be found at~\url{https://drive.google.com/file/d/1NTvo19CRzUqiUZokJaXM_4WVogXj1sdb/view?usp=sharing}. 

\section{Subjective Evaluation}

Although statistical and objective measures indicates that the model succeeds in capturing the underlying data distribution for generating novel lyrics and melodies, it is still difficult to conclude that the automatically generated composition pleases human ears. Further, lyric writing and melody composition are human creative process hence, it is very challenging to precisely quantify the automatically generated lyrics and melodies objectively and statistically. Hence, we adopt the subjective evaluation method proposed in~\cite{yu2019conditional} and~\cite{lee2019icomposer} for evaluating generated lyrics and melodies by our AutoNLMC. We ask the following questions to participants during subjective evaluation 
\begin{itemize}
    \item How meaningful are the predicted lyrics?
    \item How well does the melodies fit the lyrics?
\end{itemize}

\noindent Participants rate the given samples on a five point discrete scale from 1 to 5 (where 1 corresponds to "very bad", 2 to "bad", 3 to "ok", 4 to "good", and 5 to ""very good"). We consider five full length lyrics and melodies generated from AutoNLMC for subjective evaluation. We evaluate the generated lyrics and melodies with baseline method and the ground truth human compositions. We follow~\cite{yu2019conditional} and~\cite{lee2019icomposer} to create baseline lyrics and melodies. 
The attributes of the baseline melodies i.e., pitch, duration and rest are sampled from the respective data distributions except for pitch whose MIDI values sampled between 55-80, as most of the pitch distribution is concentrated in this range in our dataset. The baseline lyrics are sampled from the most frequent syllable lyrics vocabulary. We separately conduct the subjective evaluation for generated lyrics and lyric-melody pairs. The subjective evaluation is conducted through the Google sheets where Google sheets consists of the samples understudy, radio buttons indicating scores to choose from, and clear instructions on how to rate the samples based on 5 point scale. An example evaluation sheet used for lyrics evaluation can be accessed from here~\url{https://forms.gle/vZ6KnPDR6FgStizG9}. We requested 20 adults who had at least basic knowledge about lyrics, melodies and specifically about the music composition for subjective evaluation. Out of 20 request participants, we have obtained response from about 11 participants over the two weeks survey period. We randomize 5 full length generated lyrics from the proposed AutoNLMC, baseline and the ground truth lyrics. We ask the subjects to rate for the question "how meaningful are the presented lyrics". Similarly, for lyrics-melody pairs, we randomly choose lyrics-melody pairs from AutoNLMC, baseline and the ground truth human compositions. We ask the raters to rate for the question "how well does the melodies fit the lyrics" on a five point scale. The subjective evaluation results for ASW and CSWP for $\tau = 0.6$ are shown in Fig.~\ref{fig:subjective_score}. We can see that ASW is marginally better than CSWP. We can also observe that AutoNLMC is close to the human compositions. The baseline performs worst than the other methods without any surprise. We also see that the ASW performs better than CSWP which is in correlation with the objective measures of Section \ref{sec:experiments}. From the subjective evaluation measures, we can find that there is a gap between the human compositions and generated lyrics and melodies by the proposed model. Which indicated that there is a lot of scope to explore injecting prior musical knowledge to improve the current model.

\section{Summary} 
\label{sec:sum_conc}

In this paper, we proposed Automatic Neural Lyrics and Melody Composition (AutoNLMC) to enable the human community to discover the original lyrics, and the corresponding matching melodies. We trained lyrics to vector (lyric2vec) models on a large set of lyrics-melody parallel dataset parsed at syllable, word and sentence level to extract the lyric embeddings. The proposed AutoNLMC is a encoder-decoder sequential recurrent neural network model consisted of lyric generator, lyrics encoder and melody decoder for each note attribute trained end-to-end to learn the correlation between the lyrics and melody pairs with attention mechanism. AutoNLMC is designed to operate in two modes such that it can generate both lyrics and corresponding melody automatically for an amateur or person with no music knowledge, or it can take seed lyrics from professional lyric writer to generate the corresponding melodies. The qualitative and quantitative evaluation measures reveled that the proposed method indeed capable of generating original lyrics and corresponding melody for generating new songs. 

{\small
\bibliographystyle{IEEEtran}
\bibliography{mybib}
}

\end{document}